\documentclass[floatfix,aps,prd,superscriptaddress,noeprint,preprint,onecolumn,preprintnumbers]{revtex4-1}
\usepackage{amsfonts}                            % Use AMS fonts
\usepackage{amssymb}                             % and symbols
\usepackage{amsmath}  
\usepackage{mathtools}                           % and definitions
\usepackage{bm}
\usepackage{graphicx}                            % Graphics package
\usepackage{color}
\usepackage{latexsym}                            % Load additional symbols
\usepackage{dcolumn}                             % Decimal-dot aligned columns
\usepackage[colorlinks=true,linkcolor=blue,citecolor=blue,urlcolor=blue]{hyperref}
\usepackage{footnote}
\usepackage[normalem]{ulem} % changed by Peter to use normalem
\usepackage{slashed}
\usepackage{xfrac}
\usepackage{cleveref}
\crefname{figure}{Fig.}{Figs.}
\crefname{equation}{Eq.}{Eqs.}
\usepackage{natbib}
\usepackage{mathrsfs}
\usepackage{physics} 
\usepackage{comment}
\usepackage[caption=false]{subfig}
\newcommand\be{\begin{equation}}
\newcommand\ee{\end{equation}}
\newcommand\bea{\begin{eqnarray}}
\newcommand\eea{\end{eqnarray}}
\newcommand\bal{\begin{aligned}}
\newcommand\eal{\end{aligned}}
\newcommand\bes{\begin{subequations}}
\newcommand\ees{\end{subequations}}

\renewcommand{\Im}{\operatorname{Im}}
\renewcommand{\Tr}{\operatorname{Tr}}

\begin{document}
\title{Normal single-spin asymmetries in electron-proton scattering: two-photon exchange with intermediate state resonances}

\preprint{JLAB-THY-23-3835, ADP-23-15/T1224}

\author{Jaseer Ahmed}
\affiliation{\mbox{Department of Physics,
  Shahjalal University of Science and Technology}, Sylhet-3114, Bangladesh}
\author{P.~G.~Blunden}
\affiliation{\mbox{Department of Physics and Astronomy,
	University of Manitoba}, Winnipeg, Manitoba, Canada R3T 2N2}
\author{W.~Melnitchouk}
\affiliation{Jefferson Lab, Newport News, Virginia 23606, USA}
\affiliation{\mbox{CSSM and CDMPP, Department of Physics, University of Adelaide, Adelaide 5000, Australia}}

\begin{abstract} 
We calculate the beam ($B_n$) and target ($A_n$) normal single-spin asymmetries in electron--proton elastic scattering from two-photon exchange amplitudes with resonance intermediate states of spin-parity $1/2^\pm$ and $3/2^\pm$ and mass $W \lesssim 1.8$~GeV.
The latest CLAS exclusive meson electroproduction data are used as input for the transition amplitudes from the proton to the excited resonance states.
For $B_n$, the spin 3/2 resonances dominate by an order of magnitude over the spin 1/2 states. In general we observe cancellations between the negative contributions of the $\Delta(1232)$ and $N(1520)$ across both beam energy and scattering angle, and the positive contributions of the $\Delta(1700)$ and $N(1720)$, leading to a rather large overall uncertainty band in the total $B_n$. At forward angles and beam energies $E_\textrm{lab}<1$~GeV, where the $\Delta(1232)$ dominates, the calculated $B_n$ tend to overshoot the A4 and SAMPLE data.
The calculated $B_n$ compare well with the measured values from the  A4 and $Q_{\textrm{weak}}$ experiments with $E_\textrm{lab}>1$~GeV.
\end{abstract}

\date{\today}
\maketitle

%%%%%%%%%%%%%%%%%%%%%%%%%%%%%%%%%%%%%%%%%%%%%%%%%%%%%%%%%%%%%%%%%%%%%%%%%%%%%%%%%%
\section{Introduction}
\label{ssec.intro}

Over the last two decades the role of two-photon exchange (TPE) in electron--proton elastic scattering has received considerable attention in both the theoretical and experimental nuclear physics communities, in an effort to understand its impact on hadron structure dependent observables~\cite{Carlson:2007sp, Arrington:2011dn, Blunden:2017nby}.
Analysis of the proton's electric ($G_E$) to magnetic ($G_M$) form factor ratio, $\mu_p G_E/ G_M$, where $\mu_p$ is the proton's magnetic moment, extracted from both the Rosenbluth separation~\cite{Rosenbluth:1950yq} and polarization transfer methods~\cite{Ahmed:2020uso, Blunden:2005ew}, suggests a consistent description is possible with the inclusion of TPE effects, which have been found to make large contributions to the former~\cite{Guichon:2003qm, Blunden:2003sp}.
Subsequently, there has been a greater appreciation of the potential effects on other hadronic observables in electromagnetic reactions that may be affected by TPE, and particularly the careful propagation of its uncertainty~\cite{Carlson:2007sp, Arrington:2011dn, Blunden:2017nby}.

While the real part of the TPE amplitude can be accessed directly from the measurement of the ratio of the unpolarized $e^+p$ to $e^-p$ scattering cross sections, the imaginary part of TPE generates a single-spin asymmetry (SSA) at leading order in the electromagnetic coupling~$\alpha$, with either the beam or target polarized normal (or transverse) to the scattering plane.
Explicitly, the experimentally measured asymmetry is defined by
\be
\textrm{SSA} = \dfrac{\sigma^{\uparrow} - \sigma^{\downarrow}}{\sigma^{\uparrow} + \sigma^{\downarrow}},
\label{eq.SSA1}
\ee
where $\sigma^{\uparrow}~(\sigma^{\downarrow})$ is the cross section for $ep$ elastic scattering with either beam or target spin polarized parallel (antiparallel) to the scattering plane.
The normal vector $\bm{N}$ is defined as
\be
\bm{N} = \dfrac{\bm{k} \times \bm{k}'}{\vert \bm{k} \times \bm{k}'\vert},
\label{eq.SSAnormal}
\ee
where $\bm{k}$ and $\bm{k}'$ are the three-momenta of the incident and scattered electrons, respectively.
The leading term of the SSA comes from the imaginary part of the TPE amplitude.
It was first shown by de R\'ujula {\it et al.}~\cite{DeRujula:1971nnp} that time-reversal invariance implies no contribution to SSA from the single-photon exchange transition amplitude, ${\cal M}_\gamma$.
The leading term of the beam or target normal SSA arises from the absorptive part of the TPE transition amplitude ${\cal M}_{\gamma\gamma}$, denoted Abs\,$[{\cal M}_{\gamma\gamma}]$, according to the relation
\be
\textrm{SSA} 
= \dfrac{\Im \Big( \sum\limits_{\rm spins}^{} {\cal M}_\gamma^{*}\, {\rm Abs}\,[{\cal M}_{\gamma\gamma}] \Big)}
{ \sum\limits_{\textrm{spins}}^{} \vert {\cal M}_\gamma \vert^2}.
\label{eq.SSA2}
\ee
While there is some inconsistency with the notation used for this observable in the literature, in this work the convention $A_n$ for target normal SSA and $B_n$ for beam normal SSA will be used.

As defined in Eq.~(\ref{eq.SSA2}), the SSA is of order $\alpha$. 
The beam normal asymmetry $B_n$ is further suppressed by the small factor $m_e/E_\textrm{lab}$, where $m_e$ is the electron mass and $E_\textrm{lab}$ is the beam energy in the laboratory frame, so that $B_n$ is expected to be of order $10^{-6}\,\mbox{--}\,10^{-5}$ for beam energies in the GeV range.
For the target normal SSA $A_n$, on the other hand, there is no additional suppression, and hence it is anticipated to be of order $10^{-3}\,\mbox{--}\,10^{-2}$ for the same beam energy.
In addition to providing an avenue to the exploration of TPE effects, the beam normal SSA $B_n$ plays a particularly important role in parity-violating electron scattering experiments that use longitudinally polarized lepton beams to measure the asymmetry due to the spin flip. 
A nonzero $B_n$, even if small numerically, could contribute to a false asymmetry due to a slow drift in the rapid flip of the beam polarization.
As a requirement to control possible systematic errors, parity-violating experiments typically determine the beam normal SSA as a by-product.
For example, the highly precise Q$_{\rm weak}$ experiment~\cite{QWeak:2020fih} at Jefferson Lab recently determined the weak charge of the proton in a search for physics beyond the standard model, which required knowledge of the systematic error from  $B_n$ at forward scattering angles.
Several earlier parity-violating experiments~\cite{G0:2011chs, G0:2007mbp, HAPPEX:2012fud}, as well as the more recent intermediate and backward angle measurements from the A4 collaboration at Mainz~\cite{Maas:2004pd, Rios:2017vsw, Gou:2020viq}, have also determined the beam normal SSA over a range of scattering angles and energies.

Following the initial measurement by the SAMPLE collaboration~\cite{SAMPLE:2000hoe} at a beam energy $E_\textrm{lab}=0.2$~GeV and backward laboratory scattering angle, several subsequent experiments from the G0~\cite{G0:2011chs, G0:2007mbp}, HAPPEX~\cite{HAPPEX:2012fud} and Q$_{\rm weak}$~\cite{QWeak:2020fih} collaborations at Jefferson Lab and A4 at Mainz~\cite{Maas:2004pd, Rios:2017vsw, Gou:2020viq} measured $B_n$ over a wide range of scattering angles.
A trend observed in the data is the suppression of $B_n$ with increasing energy, although the correlation between energy and scattering angle is less clear. 
For backward scattering at relatively low energies, Refs.~\cite{G0:2011chs, Rios:2017vsw} find $B_n$ to be of order $\sim 10^{-5}$,
whereas the more recent measurement~\cite{Gou:2020viq} at intermediate scattering angles finds $B_n$ of order $10^{-6}$ over a similar range of beam energies.
Note that the SAMPLE~\cite{SAMPLE:2000hoe} result is in relative tension with the two other lower energy and backward angle measurements from G0~\cite{G0:2011chs} and A4~\cite{Rios:2017vsw}, which may be related to the more restricted mass range of resonance states that can contribute at the lower SAMPLE energy.
In contrast, the relatively higher energy ($1 \lesssim E_{\rm lab} \lesssim 3$~GeV) experiments~\cite{QWeak:2020fih, HAPPEX:2012fud, G0:2007mbp, Gou:2020viq} correspond to small scattering angles (with the exception of the single datum of Ref.~\cite{Gou:2020viq}), and are consistently in the range of $\approx -7$ to $-4$~ppm.

In theoretical developments, following de R\'ujula {\it et al.}~\cite{DeRujula:1971nnp} several model estimates of $B_n$ have been made in the literature~\cite{Pasquini:2004pv, Pasquini:2005yh, Gorchtein:2004ac, Gorchtein:2005yz}.
The hadronic approximation with a doubly-virtual Compton scattering analogy of the imaginary part of the TPE correction was used by Pasquini and Vanderhaeghen~\cite{Pasquini:2004pv, Pasquini:2005yh}, in which the $\pi N$ intermediate state was considered along with the elastic nucleon, with input taken from the MAID electroproduction amplitudes~\cite{Drechsel:1998hk}.
However, the model is believed to be appropriate only for forward angles.
Using a generalized parton distribution approach that is more applicable at high $Q^2$, with a real Compton scattering (RCS) analogy suitable for forward angles, Gorchtein~\cite{Gorchtein:2004ac} found rather different results, with even an opposite sign, compared to Refs.~\cite{Pasquini:2004pv, Pasquini:2005yh}.
Subsequently, Gorchtein~\cite{Gorchtein:2004ac} used a quasi-real Compton scattering (QRCS) formalism, which is more appropriate for backward angles, to estimate both $B_n$ and $A_n$, although the results are still not in agreement with that of Refs.~\cite{Pasquini:2004pv, Pasquini:2005yh}.
The significant disagreement between the measured value of beam normal SSA by the PREX collaboration~\cite{HAPPEX:2012fud} and the corresponding theoretical estimate~\cite{Gorchtein:2008dy} for heavier target nucleus $^{208}$Pb raised questions about the calculations in general.
More recently, Koshchii {\it et al.}~\cite{Koshchii:2021mqq} calculated $B_n$ for electron scattering from several spin 0 nuclei, accounting for inelastic intermediate state contributions, in addition to several other improvements on the uncertainty calculation. 
However, the result does not resolve the discrepancy between the theoretical estimates and the PREX~\cite{HAPPEX:2012fud} data for a $^{208}$Pb target nucleus. 

In contrast to the beam normal asymmetry, for the target normal SSA, $A_n$, there are currently no available data for a proton target.
An experiment to measure $A_n$ in both $e^- p$ and $e^+ p$ scattering has been proposed at Jefferson~Lab Hall~A for beam energies $E_{\rm lab} = 2.2, 4.4$ and 6.6~GeV using the Super Big-Bite Spectrometer~\cite{Grauvogel:2021btg}.
Earlier, a nonzero value of $A_n$ was found for the neutron, extracted from quasielastic electron scattering from $^3$He~\cite{Zhang:2015kna}, assuming the proton $A_n$ is given by the TPE contribution with a nucleon intermediate state~\cite{Afanasev:2002gr}.

To better understand both the beam and target normal SSAs originating from the spin-parity $1/2^{\pm}$ and $3/2^{\pm}$ resonance intermediate states associated with $\pi N$ and $\pi \pi N$ channels, we revisit the imaginary part of the TPE amplitude in $ep$ elastic scattering using the latest results for the electrocouplings extracted from recent CLAS data~\cite{HillerBlin:2019jgp, Mokeev:2008iw, CLAS:2012wxw}.
We begin in Sec.~\ref{sec.epscatt} by reviewing the kinematics of electron-proton scattering at the one- and two-photon exchange level.
In Sec.~\ref{sec.SSAgen} we introduce the single-spin asymmetries for both beam and target polarization normal to the electron-proton scattering plane and discuss the calculation of spin 1/2 and spin 3/2 intermediate state contributions.
Numerical results for the beam SSAs $B_n$ and the target SSAs $A_n$ are presented in Sec.~\ref{sec.ImTPEResult}, including a discussion of uncertainties and comparisons with available data.
A parity-violating transverse beam asymmetry, which we denote as $B_x$, can also arise from a transverse spin polarization in the scattering plane. As discussed in Sec.~\ref{sec.ImTPEResult}, this turns out to be negligibly small at the kinematics of interest.
Finally, in Sec.~\ref{sec.conclusions} we conclude with a summary of the main results of this analysis, and some discussion about future extensions of this work.

%%%%%%%%%%%%%%%%%%%%%%%%%%%%%%%%%%%%%%%%%%%%%%%%%%%%%%%%%%%%%%%%%%%%%%%%%%
\section{ELastic electron-proton scattering}
\label{sec.epscatt}

In this section, we define the general kinematic quantities needed for describing elastic electron-proton scattering.
For convenience, the calculation of SSA quantities is performed in the center-of-mass (CM) frame, although the experimental kinematics are typically given in the laboratory (or target rest) frame.
Where appropriate, we give the relevant expressions in both frames.

% ...............................................................................
\subsection{Kinematics and definitions}
\label{ssec.kinematics}

\begin{figure}[t]
\graphicspath{{ImagesK/}}
\centering
\includegraphics[width=0.8\textwidth]{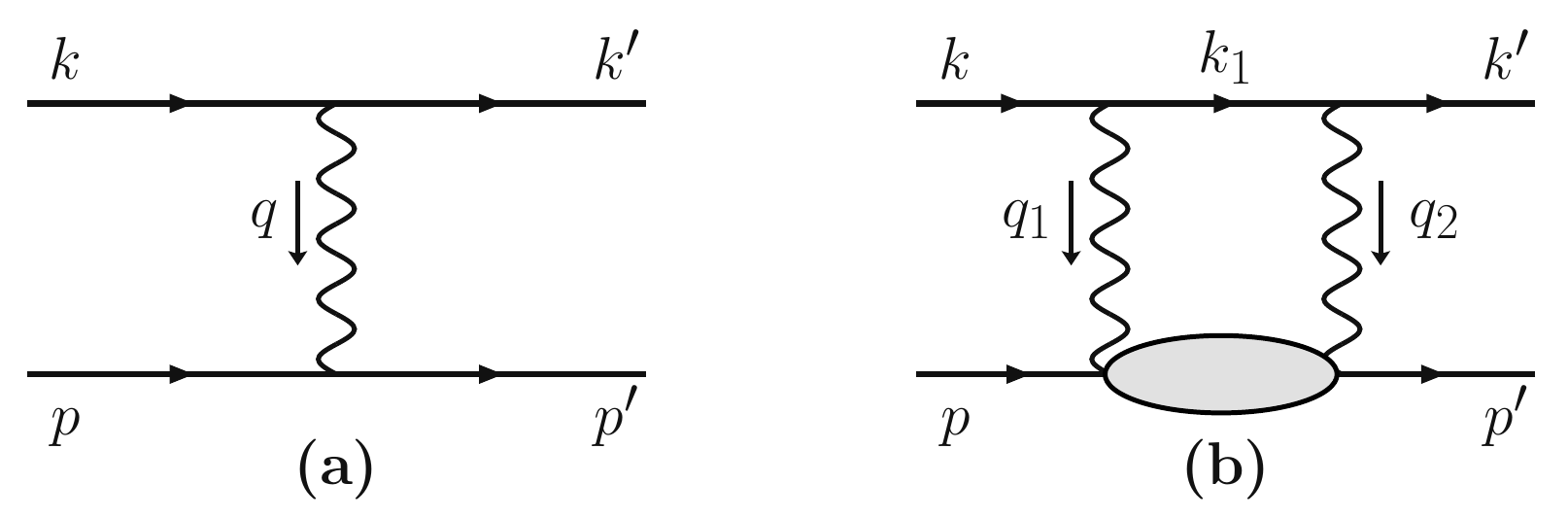}
\caption{Contributions to elastic electron--nucleon scattering from (a) one-photon exchange (OPE), and (b) two-photon exchange amplitudes, with particle momenta as indicated. The intermediate hadronic state is taken to be a resonance of invariant mass $W$. Only the $s$-channel box diagram is shown for the TPE process, since the imaginary part relevant for SSA originates solely from the box diagram, with the intermediate electron and hadronic states on-shell. The two virtual photons carry momenta $q_1$ and $q_2$, giving the total momentum transfer $q = q_1 + q_2$.}
\label{fig.tpe}
\end{figure}

For the elastic scattering process $e(k) + N(p) \to e(k') + N(p')$ (see Fig.~\ref{fig.tpe}), the four-momenta of the initial and final electrons (with mass $m_e$) are labelled by $k$ and $k'$, with corresponding lab frame energies $E_\textrm{lab}$ and $E_\textrm{lab}'$.
The initial and final nucleons (mass $M$) have four-momenta $p$ and $p'$, respectively.
The four-momentum transfer from the electron to the nucleon is given by $q = p' - p = k - k'$, with the photon virtuality $Q^2 \equiv -q^2 > 0$.
For the TPE process, the two virtual photons transfer four-momenta $q_1$ and $q_2$ to the proton, so that $q = q_1 + q_2$.

One can express the elastic scattering cross section in terms of any two of the Mandelstam invariants $s$ (total electron--nucleon invariant mass squared), $t$, and $u$, where
\be
s=(k+p)^2=(k'+p')^2\,,\quad\ \ \ \
t=(k-k')^2=q^2\,,\quad\ \ \ \
u=(p-k')^2=(p'-k)^2\,,
\label{eq:Mandelstam}
\ee
with the constraint $s + t + u = 2 M^2 + 2 m_e^2$. 
For the OPE amplitude, and for the $A_n$ SSA, the electron mass can be neglected at the kinematics of interest.
However, for the $B_n$ SSA the electron mass must be retained for two reasons.
First, $B_n$ has an overall factor of~$m_e$, and second, $B_n$ has a mass-dependent quasi-singularity when the intermediate electron three-momentum $|\bm{k}_1|\to 0$.

For the imaginary part of the scattering amplitude, the intermediate states are on-shell.
In the CM frame we have for the energies and three-momenta of the particles,
\bes
\bea
E_k &=& \frac{s-M^2+m_e^2}{2 \sqrt{s}}, \qquad |\bm{p}| = |\bm{k}| = \sqrt{E_k^2-m_e^2},
\\
E_{k_1} &=& \frac{s-W^2+m_e^2}{2 \sqrt{s}}, \qquad |\bm{k}_1| = \sqrt{E_{k_1}^2-m_e^2},
\label{eq:Ek1CM}
\\
E_p &=& \sqrt{E_k^2-m_e^2+M^2},
\eea
\label{eq:EkCM}%
\ees
where $W^2 = (p+q_1)^2 = (p'-q_2)^2$ is the invariant squared mass of the intermediate state resonance.
For the four-momentum transfer squared between the electron and nucleon, $Q^2$, and the virtualities of the two exchanged photons, $Q_1^2 = -q_1^2$ and $Q_2^2 = -q_2^2$, we have
\bes
\bea
Q^2   &=& 2 |\bm{k}|^2 \left( 1-\cos\theta_{\rm cm} \right),\\
Q_1^2 &=& 2 \left( E_k E_{k_1} - m_e^2 - |\bm{k}| |\bm{k}_1|\,\cos\theta_1 \right),\\
Q_2^2 &=& 2 \left( E_k E_{k_1} - m_e^2 - |\bm{k}| |\bm{k}_1|\,\cos\theta_2 \right),
\eea
\label{eq:Q2Q12Q22}%
\ees
where $\theta_{\rm cm}$ is the CM scattering angle, and 
$\cos\theta_2 = \cos\theta_{\rm cm} \cos\theta_1 + \sin\theta_{\rm cm} \sin\theta_1 \cos\phi_1$. 
The Mandelstam variable $s$ is given in the lab frame as $s=M^2+m_e^2 + 2 M E_{\rm lab}$, with $E_{\rm lab}$ the electron beam energy in the lab frame.
In the lab frame we also have
\bea
Q^2 = 2 E_\textrm{lab} E_\textrm{lab}' (1-\cos{\theta}_{\rm lab}),
\qquad 
E_\textrm{lab}'=E_\textrm{lab}-\dfrac{Q^2}{2 M},
\eea
where $E_\textrm{lab}'$ is the energy of the electron scattered by angle $\theta_{\rm lab}$.

For inelastic excitations the minimum value of $W$ is taken to be the pion production threshold, $W_\textrm{th}=M+m_\pi$.
For a given $s$, the maximum value of $W$ corresponds to an intermediate electron at rest, $|\bm{k}_1|=0$, so that
\bea
W_\textrm{max}=\sqrt{s}-m_e,\qquad E_{k_1}= m_e.
\label{eq:Wmax}
\eea
At $W=W_\textrm{max}$ the four-momentum transfers of the two virtual photons become
\bea
Q_1^2\ =\ Q_2^2\ =\ m_e\, \frac{\left(W_\textrm{max}^2-M^2\right)}{\sqrt{s}},
\label{eq:Q12max}
\eea
so that the two photons are almost on-shell ({\em i.e.}~real).
This has been dubbed the quasi-real Compton scattering (QRCS) region~\cite{Gorchtein:2004ac, Pasquini:2004pv, Pasquini:2005yh}, and requires special attention to reliably compute the SSA numerically.  
We discuss this further in Sec.~\ref{ssec:QRCS} and in the Appendix.

% ...............................................................................
\subsection{One- and two-photon exchange amplitudes}

The explicit expression for the one-photon exchange (OPE) or Born amplitude, ${\cal M}_{\gamma}$, of Fig.~\ref{fig.tpe} can be written as~\cite{Arrington:2011dn}
\be
{\cal M}_\gamma = e^2\, \bar{u}_e(k')\, \gamma_\mu\, u_e(k)\, \dfrac{1}{Q^2}\, \bar{u}_N(p')\, \Gamma^\mu(q)\,
u_N(p),
\label{eq.MgammaSSA}
\ee
where $e$ is the charge of the proton, and the hadronic current operator $\Gamma^\mu$ is parameterized in terms of the Dirac $F_1$ and Pauli $F_2$ form factors for on-shell particles,
\be
\Gamma^{\mu}(q) = F_1(Q^2)\gamma^{\mu} + F_2(Q^2)\dfrac{\textit{i}
\sigma^{\mu\nu} q_{\nu}}{2M}.
\label{Mgamma2}
\ee 

The two-photon exchange amplitude, ${\cal M}_{\gamma \gamma}$, contains contributions from the box diagram of Fig.~\ref{fig.tpe} and the corresponding crossed-box diagram (not shown).
However, since the crossed-box amplitude is purely real, we will focus only on the box diagram contribution, ${\cal M}_{\gamma \gamma}^{\textrm{box}}$.
The loop integral of the box diagram amplitude can be written as~\cite{Arrington:2011dn}, %
\be
{\cal M}_{\gamma \gamma}^{\textrm{box}}
= -i e^4 \int\!\dfrac{\dd^4 q_1}{(2\pi)^4}
    \dfrac{L_{\mu\nu} H^{\mu\nu}}{(q_{1}^2-\lambda^2)(q_2^2-\lambda^2)},
\label{eq.tpeBox}
\ee
where $\lambda$ is an infinitesimal photon mass introduced to regulate infrared divergences.
Such divergences are absent for normal single-spin asymmetries, but $\lambda$ can be kept as an infinitesimal parameter to improve numerical stability.
The leptonic and hadronic tensors, $L_{\mu\nu}$ and $H^{\mu\nu}$, respectively, are given by
\bes
\bea
L_{\mu\nu} &=& \bar{u}_e(k')\, \gamma_{\mu} \dfrac{(\slashed{k}_1 + m_e)}{k_1^2 - m_e^2 + i 0^+}\, \gamma_{\nu}\, u_e(k),
\\[0.2cm]
H^{\mu\nu} &=& \bar{u}_N(p')\,
  \Gamma_{R\to \gamma N}^{\mu\alpha}(p_R,-q_2)\,
  S_{\alpha\beta}(p_R,W)\,
  \Gamma_{\gamma N\to R}^{\beta\nu}(p_R,q_1)\,
  u_N(p),
  \label{eq.LHtensor}
\eea
\ees
where the intermediate lepton four-momentum is $k_1=k-q_{1}$, and the four-momentum of the resonance $R$ (with mass $W$) is $p_R = p + q_1 = p' - q_2$.
The transition operators, $\Gamma_{\gamma N \to R}^{\beta\nu}(p_R, q_1)$ and $\Gamma_{R \to \gamma N}^{\mu\alpha}(p_R,-q_2)$, between the nucleon and intermediate state resonance~$R$ can be expressed in terms of the three transition form factors $G_1$, $G_2$, and $G_3$.
These form factors can also be written in terms of the corresponding helicity amplitudes $A_{1/2}$, $A_{3/2}$, and $S_{1/2}$ (see Ref.~\cite{Ahmed:2020uso}).

For spin 1/2 baryon intermediate states, the propagator $S_{\alpha\beta}(p_R,W)$ is simply the spin 1/2 Feynman propagator,
\be
S_{\alpha \beta}(p_{R},W)
= \delta_{\alpha\beta}\,
  \dfrac{(\slashed{p}_{R}+W)}{p_{R}^2-W^2+i 0^+}
= \delta_{\alpha\beta}\, S_F(p_{R},W).
\ee
For spin 3/2 intermediate states, on the other hand, the hadronic propagator has the more complicated form
\be
S_{\alpha\beta}(p_R, W)
= -{\cal P}_{\alpha\beta}^{3/2}(p_R)\,
  \dfrac{(\slashed{p}_R + W)}{p_R^2 - W^2 + i 0^+},
\ee
where
\be
{\cal P}_{\alpha\beta}^{3/2}(p_R)
= g_{\alpha\beta}
- \dfrac{1}{3} \gamma_\alpha \gamma_\beta 
- \dfrac{1}{3 p_R^2}
  \left( \slashed{p}_R \gamma_\alpha (p_R)_\beta
        + (p_R)_\alpha \gamma_\beta \slashed{p}_R
  \right)
\ee
is the spin 3/2 projection operator for momentum $p_R$.

%%%%%%%%%%%%%%%%%%%%%%%%%%%%%%%%%%%%%%%%%%%%%%%%%%%%%%%%%%%%
\section{Single-spin asymmetries in electron-proton scattering}
\label{sec.SSAgen}

In this section we discuss several technical aspects of the TPE amplitude, including the generalization of the calculation from point particles to the case of finite resonance widths (Sec.~\ref{ssec:finitewidth}), and the quasi-singular behavior of the asymmetry $B_n$ (Sec.~\ref{ssec:QRCS}).
We begin, however, with some general considerations about TPE amplitudes and their contributions to SSAs.

% ..........................................................................
\subsection{General features}

In the definition of the beam or target normal SSA in Eq.~(\ref{eq.SSA2}), the denominator is identical to the Born cross section for unpolarized elastic $ep$ scattering, since the spin components (beam or target) have no impact at the Born level.
Summing over final state spins and averaging over initial state spins, one can write the squared Born amplitude in terms of the invariant Mandelstam variables $s$ and $Q^2=-t$,
\be
\sum\limits_{\textrm{spins}}^{} \big| {\cal M}_\gamma \big|^2 
= \sum\limits_{\textrm{spins}}^{} {\cal M}_\gamma^\dagger {\cal M}_\gamma 
= \dfrac{Q^4}{e^4}\,D(s,Q^2),
\label{eq.MgammaDsQ}
\ee
where we define the factor
\bea
D(s,Q^2) 
&=& \frac{2}{4 M^2+Q^2}
\Big[ G_E^2(Q^2)\, 8 M^2 \big( (s-M^2)^2 - Q^2 s \big)
\nonumber\\
& & \hspace*{2cm}
+\, G_M^2(Q^2) Q^2 \big( 2 M^4+Q^4+2(s-2 M^2)(s-Q^2) \big)
\Big],
\eea
with terms of order of $m_e^2$ neglected. 

To derive the absorptive part of the TPE amplitude, one can exploit the Cutkosky cutting rules~\cite{Cutkosky:1960sp}, which involve the replacements
\begin{subequations}
\bea
\frac{1}{p_R^2-W^2+i0^+}\
&\to\ & -2\pi i\, \theta(p_R^0)\, \delta({p_R^2-W^2}), 
\\
\frac{1}{k_1^2-m_e^2+i0^+}\
&\to\ & -2\pi i\, \theta(k_1^0)\, \delta({k_1^2-m_e^2}),
\eea
\end{subequations}
which place the intermediate state lepton and hadron on their mass-shells.
This substitution provides the discontinuity, 
    ${\rm Disc}(i{\cal M}_{\gamma\gamma}) = -2\Im{{\cal M}_{\gamma\gamma}}$,
of the TPE box diagram of Fig.~\ref{fig.tpe}, and hence the absorptive part of TPE amplitude $-{\rm Abs}\, {\cal M}_{\gamma\gamma}$. 
After applying the Cutkosky cutting rules, the absorptive part of the TPE amplitude in Eq.~(\ref{eq.SSA2}) can be written as
\be
{\rm Abs}\, {\cal M}_{\gamma\gamma}
= e^4 \int\!\dfrac{\dd^3 \bm{k}_1}{(2\pi)^3 2 E_{k_1}}
    \dfrac{\bar{u}_e(k') \gamma_{\mu} (\slashed{k}_1 + m_e) \gamma_{\nu} u_e(k)}{Q_1^2\, Q_2^2}\, {\cal W}^{\mu\nu}.
\label{eq.AbsM2gamma}    
\ee
The hadronic tensor ${\cal W}^{\mu\nu}$ in Eq.~(\ref{eq.AbsM2gamma}) contains all the information about the transition from the proton initial state to all possible intermediate hadronic states, including the elastic nucleon state and the inelastic transitions to the nucleon excited state resonances.
In practice, the SSAs are calculated including contributions from each of the three-star and four-star, spin 1/2 and 3/2 resonance intermediate states from the PDG~\cite{ParticleDataGroup:2018ovx} below mass $M_R = 1.8$~GeV, which are then added together with the elastic nucleon contribution to obtain the complete result.

In the zero width approximation, for the elastic nucleon and inelastic spin 1/2 resonances of mass $M_R$ the hadronic tensor ${\cal W}^{\mu\nu}$ takes the simplified form,
\be
{\cal W}^{\mu\nu}\,
=\, 2\pi \delta(W^2 - M_R^2)\, \bar{u}_N(p')\,
   \Gamma_{R \to \gamma N}^\mu(p_R, -q_2)\,
   (\slashed{p}_R + W)\,
   \Gamma_{\gamma N \to R}^\nu(p_R, q_1)\, 
   u_N(p).
\ee
To assess the validity of this approximation, we will also examine the effect of replacing the zero width result by a finite width distribution in $W^2$, centred around $W=M_R$.
For spin 3/2 resonances, the hadronic tensor uses the Rarita-Schwinger spinors for each intermediate state, and can be written as
\bea
{\cal W}^{\mu\nu}
&=& -2\pi \delta(W^2 - M_R^2)
\nonumber\\
&& \times\, \bar{u}_N(p')\,
    \Gamma_{R \to \gamma N}^{\mu\alpha}(p_R, -q_2)\,
    {\cal P}_{\alpha\beta}^{3/2}(p_R)\,(\slashed{p}_R + W)\,
    \Gamma_{\gamma N \to R}^{\beta\nu}(p_R, q_1)\, u_N(p).
\eea

Using Eqs.~(\ref{eq.MgammaSSA}), (\ref{eq.MgammaDsQ}), and (\ref{eq.AbsM2gamma}) one can write the SSA as
\bea
\textrm{SSA} &=& \dfrac{\alpha Q^2}{2 \pi^2 D(s,Q^2)}
\nonumber\\
&&\times \sum\limits_{\textrm{spins}}^{}\int\, \dfrac{\dd^3\bm{k}_1}{2 E_{k_1}}\,\dfrac{\bar{u}_e(k)\, \gamma_\rho\, u_e(k')\, \bar{u}_e(k')\, \gamma_{\mu} (\slashed{k}_1 + m_e) \gamma_{\nu}\, u_e(k)\, \bar{u}_N(p)\, \Gamma^\rho(-q)\, u_N(p')}{Q_1^2\, Q_2^2}\, {\cal W}^{\mu\nu}.
\nonumber\\
&&
\label{eq.SSA4}
\eea
For the two different cases of beam and target normal SSA, the spin sum will lead to different expressions for the SSAs.
Taking the spin sum, one can express Eq.~(\ref{eq.SSA4}) in a concise form in terms of the leptonic and hadronic tensors, $L_{\rho\mu\nu}$ and $H^{\rho\mu\nu}$, respectively, as
\be
\textrm{SSA} = \dfrac{\alpha Q^2}{\pi D(s,Q^2)}\int\, \dfrac{\dd^3\bm{k}_1}{2 E_{k_1}}\dfrac{\Im L_{\rho\mu\nu} H^{\rho\mu\nu}}{Q_1^2\, Q_2^2}.
\label{eq.SSA5}
\ee
For the beam polarized parallel or antiparallel to the normal $\bm{s}_n$ to the scattering plane defined in Eq.~(\ref{eq.SSAnormal}), % {\em i.e.}~for $B_n$, 
the leptonic tensor $L_{\rho\mu\nu}$ contains the lepton polarization vector $s_n^\mu \equiv (0; \bm{s}_n)$, and takes the form
\be
L^\textrm{B}_{\rho\mu\nu} 
= \frac12
\Tr\big[ (1+\gamma_5 \slashed{s}_n)
        (\slashed{k} + m_e) \gamma_\rho 
        (\slashed{k}' + m_e) \gamma_\mu
        (\slashed{k}_1 + m_e) \gamma_\nu
\big],
\label{eq.LBn}
\ee
where the superscript ``B'' denotes the fact that the lepton tensor corresponds to the beam normal case. 
Note that the imaginary part in Eq.~(\ref{eq.SSA2}) for $B_n$ comes solely from this spin polarization-dependent term. 
However, the corresponding hadronic tensor for the beam normal case, $H^{\rho\mu\nu}_{\rm B}$, remains independent of the polarization of the target hadron, and is equivalent to the hadronic tensor for the case of unpolarized $ep$ scattering.

For spin 1/2 intermediate states, the hadronic tensor becomes
\bea
H^{\rho\mu\nu}_{\rm B} 
&=& \frac12 
\Tr\big[ (\slashed{p} + M)\, \Gamma_\rho(-q)\,
          (\slashed{p}' + M)\, \Gamma_{R \to \gamma N}^\mu(p_R, -q_2)
\nonumber\\
& & \hspace*{0.9cm} \times
          (\slashed{p}_R + W)\, \Gamma_{\gamma N \to R}^\nu(p_R, q_1)
\big]\,
\delta(W^2 - M_R^2).
\label{eq.HBn12}
\eea
For spin 3/2 resonances, on the other hand, the hadronic tensor is given by
\bea
H^{\rho\mu\nu}_{\rm B} 
&=& -\frac12 
\Tr\big[ (\slashed{p} + M)\, \Gamma_\rho(-q)\,
          (\slashed{p}' + M)\, \Gamma_{R \to \gamma N}^{\mu\alpha}(p_R, -q_2)\,
          {\cal P}_{\alpha\beta}^{3/2}(p_R)
\nonumber\\
& & \hspace*{1.2cm}\times
          (\slashed{p}_R + W)\, \Gamma_{\gamma N \to R}^{\beta\nu}(p_R, q_1)
\big]\,
\delta(W^2 - M_R^2).
\label{eq.HBn32}
\eea

For the target normal SSA, $A_n$, the corresponding leptonic tensor, $L^{\rm A}_{\rho\mu\nu}$, is identical to that for unpolarized $ep$ scattering, and can be written as
\be
L^\textrm{A}_{\rho\mu\nu} 
= \frac12 
\Tr\big[ (\slashed{k} + m_e)\, \gamma_\rho\,
        (\slashed{k}' + m_e)\, \gamma_\mu\,
        (\slashed{k}_1 + m_e)\, \gamma_\nu
\big].
\label{eq.LAn}
\ee
Unlike for $B_n$, the hadronic tensor $H^{\rho\mu\nu}_{\rm A}$ for the target normal SSA $A_n$ depends on the target polarization vector, $S_n^{\mu}$.
For spin 1/2 resonances, $H^{\rho\mu\nu}_{\rm A}$ becomes
\bea
H^{\rho\mu\nu}_{\rm A}
&=& \frac12 
\Tr\big[ (1+\gamma_5 \slashed{S}_n) 
        (\slashed{p} + M)\, \Gamma_\rho(-q)\,
        (\slashed{p}' + M)\, \Gamma_{R \to \gamma N}^\mu(p_R, -q_2)\,
\nonumber\\ 
&& \hspace*{0.9cm} \times
        (\slashed{p}_R + W)\, \Gamma_{\gamma N \to R}^\nu(p_R, q_1)
\big]\,
\delta(W^2 - M_R^2),
\label{eq.HAn12}
\eea
while for spin 3/2 resonances it is given by
\bea
H^{\rho\mu\nu}_{\rm A} 
&=& -\frac12
\Tr\big[ (1+\gamma_5 \slashed{S}_n)
        (\slashed{p} + M)\, \Gamma_\rho(-q)\,
        (\slashed{p}' + M)\, \Gamma_{R \to \gamma N}^{\mu\alpha}(p_R, -q_2)\,
        {\cal P}_{\alpha\beta}^{3/2}(p_R)
\nonumber\\
&& \hspace*{1.2cm} \times
        (\slashed{p}_R + W)\, \Gamma_{\gamma N \to R}^{\beta\nu}(p_R, q_1)
\big]\,
\delta(W^2 - M_R^2).
\label{eq.HAn32}
\eea

For the numerical calculation, it will be convenient to transform the phase space integral over the intermediate electron momentum $\bm{k}_1$ of Eq.~(\ref{eq.SSA5}) in terms of the Lorentz-invariant Mandelstam variable $s$. 
Defining the kinematics in the CM frame, the integration over
    $\dd^3\bm{k}_1 \to 
    \bm{k}_1^2\, \dd\vert\bm{k}_1\vert\, \dd(\cos\theta_{k_1})\, \dd\phi_{k_1}$ 
can be written as
\be
\int \frac{\dd^3\bm{k}_1}{2 E_{k_1}}\, \to\, - \int_{M^2}^{W_\textrm{max}^2} \dd W^2\  \dfrac{\vert\bm{k}_1\vert}{4\,\sqrt{s}}\int_{-1}^1 \dd\cos\theta_{k_1}\int_0^{2\pi} \dd\phi_{k_1},
\label{eq.d3k1}
\ee
with $W_\textrm{max}=\sqrt{s}-m_e$. 
Here we have utilized the CM frame relation for the intermediate electron three-momentum given in Eq.~(\ref{eq:Ek1CM}).
\be
\textrm{SSA} = -\, \dfrac{\alpha\, Q^2}{\pi D(s,Q^2)}\ 
\int_{M^2}^{W_\textrm{max}^2} \dd W^2
\ \frac{\vert\bm{k}_1\vert}{4\,\sqrt{s}} 
\int_{-1}^1 \dd\cos\theta_{k_1}
\int_0^{2\pi} \dd\phi_{k_1}\,
\dfrac{\Im L_{\rho\mu\nu} H^{\rho\mu\nu}}{Q_1^2\, Q_2^2}.
\label{eq.SSA6}
\ee
The tensor product $L_{\rho\mu\nu} H^{\rho\mu\nu}$ in Eq.~(\ref{eq.SSA6}) depends on the totally antisymmetric Levi-Civita tensor, $\epsilon_{\alpha\beta\gamma\delta}$, which is defined
following the FeynCalc~\cite{Shtabovenko:2016sxi} convention $\epsilon_{0123} = -1 = -\epsilon^{0123}$.
In the following we will use the shorthand notation 
$\epsilon(abcd)
 \equiv \epsilon_{\alpha\beta\gamma\delta}a^\alpha b^\beta c^\gamma d^\delta$.
For the beam normal spin asymmetry $B_n$ there are four independent antisymmetric tensors that be constructed from the beam normal spin four-vector $s_n$ and three of the four-momenta $k$, $p$, $q$, and $q_1$. For the target normal spin asymmetry $A_n$ there is one  antisymmetric tensor needed.
In the CM frame these can be written as
\bes
\bea
\epsilon(k p q s_n) &=& -(E_k + E_p) |\bm{k}|^2 \sin{\theta_{\rm cm}},
\\
\epsilon(k p q_1 s_n) &=& -(E_k + E_p) |\bm{k}| |\bm{k}_1| \sin{\theta_{k_1}}  \cos{\phi_{k_1}},
\\
\epsilon(k q q_1 s_n) &=& 
  |\bm{k}|\, \big(\, 
  \big[ E_k |\bm{k}_1| \cos{\theta_{k_1}} - E_{k_1} |\bm{k}| \big] 
  \sin{\theta_{\rm cm}}
  \nonumber\\
&&\hspace*{0.6cm}
+\, E_k |\bm{k}_1| (1 - \cos{\theta_{\rm cm}}) \sin{\theta_{k_1}} \cos{\phi_{k_1}} 
    \big),
\\
\epsilon(p q q_1 s_n) &=& 
  |\bm{k}|\, \big( 
  \big[ E_p |\bm{k}_1| \cos{\theta_{k_1}} - (E_k - E_{k_1} + E_p) |\bm{k}| \big]
  \sin{\theta_{\rm cm}} 
\nonumber\\
&&\hspace*{0.6cm}
+\, E_p |\bm{k}_1| (1 - \cos{\theta_{\rm cm}}) \sin{\theta_{k_1}} \cos{\phi_{k_1}} 
    \big),\\
\epsilon(k p q q_1) &=& (E_k + E_p) |\bm{k}| |\bm{k}|^2 \sin{\theta_{\rm cm}} \sin{\theta_{k_1}} \sin{\phi_{k_1}}.
\eea
\ees

% ..........................................................................
\subsection{Finite width effect}
\label{ssec:finitewidth}

A finite resonance width is usually accommodated by using the well-known relativistic Breit-Wigner distribution in $W^2$.
In this analysis we use a closely related alternative distribution, denoted as a Sill distribution by Giacosa~{\em et al.}~\cite{Giacosa:2021mbz}, that avoids the problem of normalization inherent in the Breit-Wigner expression.
In this approach the $\delta$-function distribution $\delta(W^2-M_R^2)$ that appears in Eqs.~(\ref{eq.HAn12}) and (\ref{eq.HAn32}) is replaced by the function
\bea
\delta_\textrm{Sill}(W^2)
&=&\frac{\theta(W^2-W_{\rm th}^2)}{\pi} 
\frac{\sqrt{W^2-W_{\rm th}^2}\ \widetilde\Gamma}{(W^2-M_R^2)^2 + (W^2-W_{\rm th}^2)\,{\widetilde\Gamma}^2},
\label{eq.BW}
\eea
where
\bea
{\widetilde\Gamma} 
&=& \Gamma \frac{M_R}{\sqrt{M_R^2-W_{\rm th}^2}}
\eea
and $\Gamma$ is the usual resonance width.
The Sill distribution has the desirable property that
\be
\int_{W_\textrm{th}^2}^\infty \dd W^2\, \delta_\textrm{Sill}(W^2) = 1
\ee
for any threshold $W_\textrm{th}^2 < M_R^2$.
It vanishes as $W \to W_\textrm{th}$, but is otherwise very similar to the conventional Breit-Wigner distribution.

% ..........................................................................
\subsection{Quasi-singular behavior in $B_n$}
\label{ssec:QRCS}

As discussed at the end of Sec.~\ref{ssec.kinematics}, the beam normal SSA $B_n$ is sensitive to the quasi-singular behavior of the integrand in Eq.~(\ref{eq.SSA6}) when the intermediate state electron three-momentum $|\bm{k}_1| \to 0$.
This is the QRCS region, where $W\to W_\textrm{max}$ and the two virtual photons have four-momenta $Q_1^2$ and $Q_2^2$ of order $m_e$ (see Eq.~(\ref{eq:Q12max})).
In this region of $W$, the integrand of Eq.~(\ref{eq.SSA6}) is characterized by a slowly varying numerator and a rapidly varying denominator.
This behaviour does not affect the target normal SSA $A_n$ because for this asymmetry the numerator in Eq.~(\ref{eq.SSA6}) vanishes as $W \to W_\textrm{max}$.

To address this behavior in the numerical calculations in a practical way, we have devised the following strategy in the QRCS region with $W$ just below $W_\textrm{max}$.
The slowly varying numerator of the integrand in Eq.~(\ref{eq.SSA6}) is evaluated at $Q_1^2=Q_2^2=0$, which is then a constant independent of $\theta_{k_1}$ and $\phi_{k_1}$.
We keep the mild $W$ dependence, but make no further approximation and leave the denominator intact.
Thus we are left with an integral over $W$ in this region that is proportional to the angular integral
\bea
J(W) = \frac{|\bm{k}_1|}{4\sqrt{s}} \int \dd\Omega_{k_1}\ \frac{1}{Q_1^2\,Q_2^2}.
\label{eq.JW}
\eea
This integral can be done analytically, as discussed in Refs.~\cite{Afanasev:2004pu, Gorchtein:2005yz, Blunden:2017nby}.
Unlike Refs.~\cite{Afanasev:2004pu, Gorchtein:2005yz} however, we only apply the analytic expression using $J(W)$ to the tail region, $W_\textrm{max}-5 m_e\leq W \leq W_\textrm{max}$, and use the full three-dimensional numerical quadrature of Eq.~(\ref{eq.SSA6}) elsewhere. Details of the matching procedure at $W=W_\textrm{max}-5 m_e$ and the analytic expression for $J(W)$ are given in the Appendix.

%%%%%%%%%%%%%%%%%%%%%%%%%%%%%%%%%%%%%%%%%%%%%%%%%%%%%%%%%%%%%%%%%%%%%%%%%%%
\section{Numerical single-spin asymmetry results}
\label{sec.ImTPEResult}

In this section we present the results of our calculation of single-spin asymmetries for both beam (Sec.~\ref{ssec.BnResult}) and target (Sec.~\ref{ssec.AnResult}) spin normal to the scattering plane, at the kinematics of several previous experiments.
Before discussing the results for $B_n$ and $A_n$, we will illustrate the input parameters used in the evaluation of the integral in Eq.~(\ref{eq.SSA6}). 

% ........................................................................
\subsection{Resonance parameters}
\label{ssec.RParameters}

In our numerical calculations, for the proton elastic electric ($G_E$) and magnetic ($G_M$) form factors we use the parametrization from Ref.~\cite{Arrington:2007ux}.
For the hadronic transition currents $\Gamma_{R\to \gamma N}$ and $\Gamma_{\gamma N \to R}$ in Eq.~(\ref{eq.SSA6}), we use the CLAS parametrization~\cite{HillerBlin:2019jgp} of the input resonance electrocouplings $A_h(Q^2)$ at the resonance points, where $A_h$ represents the longitudinal electrocoupling, $S_{1/2}$, and the two transverse electrocouplings, $A_{1/2}$ and $A_{3/2}$.
The dependence of the electrocouplings $A_h$ on the invariant mass $W$ is given in Ref.~\cite{Ahmed:2020uso}.

For the inelastic intermediate states in Fig.~\ref{fig.tpe}(b), in this work we include the contributions of four spin-parity $3/2^\pm$ nucleon (isospin 1/2) and $\Delta$ (isospin 3/2) resonances 
\{$\Delta(1232)\,3/2^+$, $N(1520)\,3/2^-$, $\Delta(1700)\,3/2^-$, and $N(1720)\,3/2^+$\}, and five spin-parity $1/2^\pm$ resonances
\{$N(1440)\,1/2^+$, $N(1535)\,1/2^-$, $\Delta(1620)\,1/2^-$, $N(1650)\,1/2^-$, and $N(1710)\,1/2^+$\}.
(In the following, for ease of notation we will drop the spin-parity suffix from the resonance state labels.)
The Breit-Wigner mass $M_R$ and the constant decay width $\Gamma$ of the nine excited state resonances are set to those used in the CLAS parametrization \cite{HillerBlin:2019jgp} of the resonance electrocouplings $A_h$, and their numerical values are listed in the second and third columns of Table~\ref{tab.AhUncr}.

Laboratory threshold energies $E_{\textrm{lab}}^{\textrm{th}} = (M_R^2 - M^2)/2M$ for the excitation of resonances $R$ in the zero width limit are shown in Table~\ref{tab.AhUncr}. Values range between $0.34$~GeV for the first excited state $\Delta(1232)$ to $1.11$~GeV for the highest-mass state $N(1720)$. 
It is evident from the threshold energy values that in the zero width approximation the states beyond the $N(1650)$ do not contribute to the total SSA for beam energies below 1~GeV, where most of the experiments to measure $B_n$ have taken data. 
In practice, the unstable resonances have a finite decay width with a distribution in the squared invariant mass $W^2$, starting from the threshold, $W_{\textrm{th}}^2$, of the prominent $n \pi^+$ decay channel of most resonances. 
Accounting for the finite width effect for each resonance, using the Sill distribution of Eq.~(\ref{eq.BW}), gives a nonzero contribution from the higher-mass resonances even at beam energies $E_{\textrm{lab}} \lesssim 1.0$~GeV. 
The effect of such a nonzero width on the beam and target SSAs $B_n$ and $A_n$ will be discussed in more detail below.

\begin{table}[b]
\centering
\caption{Mass ($M_R$), width ($\Gamma$), and threshold energy ($E_\textrm{lab}^\textrm{th}$) of each $N$ or $\Delta$ resonance of spin-parity $J^\pm$. The uncertainty bands $\Delta A_{1/2}$, $\Delta A_{3/2}$ and $\Delta S_{1/2}$ on the respective electrocouplings $A_{1/2}$, $A_{3/2}$ and $S_{1/2}$, used in estimating the uncertainty in $B_n$ and $A_n$, are given in the last three columns. The uncertainties $\Delta A_h$ are given as a percentage of the maximum absolute value of the corresponding electrocouplings, except for the $\Delta(1232)$, where $\Delta A_{1/2}$ and $\Delta A_{3/2}$ are given as a percentage of $A_{1/2}$ and $A_{3/2}$, respectively.}
\begin{tabular*}{\textwidth}{l @{\hspace{12pt}} @{\extracolsep{\fill}}  lcl rrr @{}}
\hline\hline
Resonance & $M_R$ (GeV) & $\Gamma$ (GeV) & $E_\textrm{lab}^\textrm{th}$ (GeV) & $\Delta A_{1/2}(\%)$ & $\Delta A_{3/2}(\%)$ & $\Delta S_{1/2}(\%)$ \\
\hline
$\Delta(1232)\,3/2^+$ & ~~1.232 & 0.117 & ~~~~0.34 
& 3.0~~~~~ & 4.5~~~~~ & 3.6~~~~~\\

$N(1440)\,1/2^+$ & ~~1.430 &0.350 &~~~~0.64 
& 10.0~~~~~ & ---~~~~~ & 15.9~~~~~\\

$N(1520)\,3/2^-$ & ~~1.515 & 0.115 & ~~~~0.75 
& 6.1~~~~~ & 5.3~~~~~ & 8.9~~~~~\\

$N(1535)\,1/2^-$ & ~~1.535&0.150& ~~~~0.78
& 5.0~~~~~ & ---~~~~~ & 22.1~~~~~\\

$\Delta(1620)\,1/2^-$ & ~~1.630 &0.140 & ~~~~0.91 
& 21.2~~~~~ & ---~~~~~ & 12.1~~~~~\\

$N(1650)\,1/2^-$ & ~~1.655 & 0.140 & ~~~~0.98 
& 15.8~~~~~ & ---~~~~~ & 23.6~~~~~\\

$\Delta(1700)\,3/2^-$ & ~~1.700& 0.293 & ~~~~1.09 
& 5.0~~~~~ & 9.1~~~~~ & 12.9~~~~~\\

$N(1710)\,1/2^+$ & ~~1.710 & 0.100 & ~~~~1.09
& 15.0~~~~~ & ---~~~~~ & 49.2~~~~~\\

$N(1720)\,3/2^+$ & ~~1.748 &0.114  & ~~~~1.11
& 4.5~~~~~ & 10.7~~~~~ & 13.8~~~~~\\
%
%$N'(1720)3/2^+$ & ~~1.725&0.12& -- & -- & -- & --\\
\hline\hline
\end{tabular*}
\label{tab.AhUncr}
\end{table}

% ........................................................................
\subsection{Uncertainty estimation}
\label{ssec.uncer}

Apart from the dependence on the width, we also propagate the uncertainty on the input resonance electrocouplings, $\Delta A_h$, into the estimation of the uncertainties on the beam normal SSA $B_n$, using
\bea
\Delta B_n = \sqrt{\Big(\dfrac{\partial B_n}{\partial A_{1/2}}\Big)^2 (\Delta A_{1/2})^2 + \Big(\dfrac{\partial B_n}{\partial A_{3/2}}\Big)^2(\Delta A_{3/2})^2 + \Big(\dfrac{\partial B_n}{\partial S_{1/2}}\Big)^2(\Delta S_{1/2})^2}\, ,
\eea
and similarly for the uncertainty, $\Delta A_n$, on the target normal asymmetry, $A_n$.
A constant, $Q^2$-independent uncertainty on the electrocouplings was assumed for each of the resonances in the range of $0 \leq Q^2 \leq 5$~GeV$^2$, with the exception of the $\Delta(1232)$, for which there is more empirical information.
The uncertainties on the transverse $A_{1/2}$ and $A_{3/2}$ electrocouplings of the $\Delta(1232)$ transition display some $Q^2$ dependence, and decrease with $Q^2$, following the magnitudes of the respective electrocouplings~\cite{HillerBlin:2019jgp}. 
As shown in Table~\ref{tab.AhUncr}, the uncertainties $\Delta A_{1/2}$ and $\Delta A_{3/2}$ on the two transverse electrocuplings are assumed to be $\approx 3\%$ and $4.5\%$ of the corresponding electrocouplings, respectively.
For the longitudinal electrocoupling $S_{1/2}$, the uncertainty $\Delta S_{1/2}$ for the $\Delta(1232)$ transition (similar to all other resonances) can be approximated by a constant $\sim 3.6\%$ of the maximum value of $S_{1/2}$ \cite{HillerBlin:2019jgp}, which occurs at $Q^2 = 0.127$~GeV$^2$. 
The constant uncertainties $\Delta{A_h}$ for the remaining states are given in Table~\ref{tab.AhUncr} as a percentage of the maximum value of the corresponding electrocouplings $A_h$ over the range $0 \leq Q^2 \leq 5$~GeV$^2$. 

% ......................................................................
\subsection{Beam normal SSA $B_n$}
\label{ssec.BnResult}

\begin{figure}[ht!]
\graphicspath{{ImagesK/}}
\centering
\subfloat{
\includegraphics[width=0.49\textwidth]{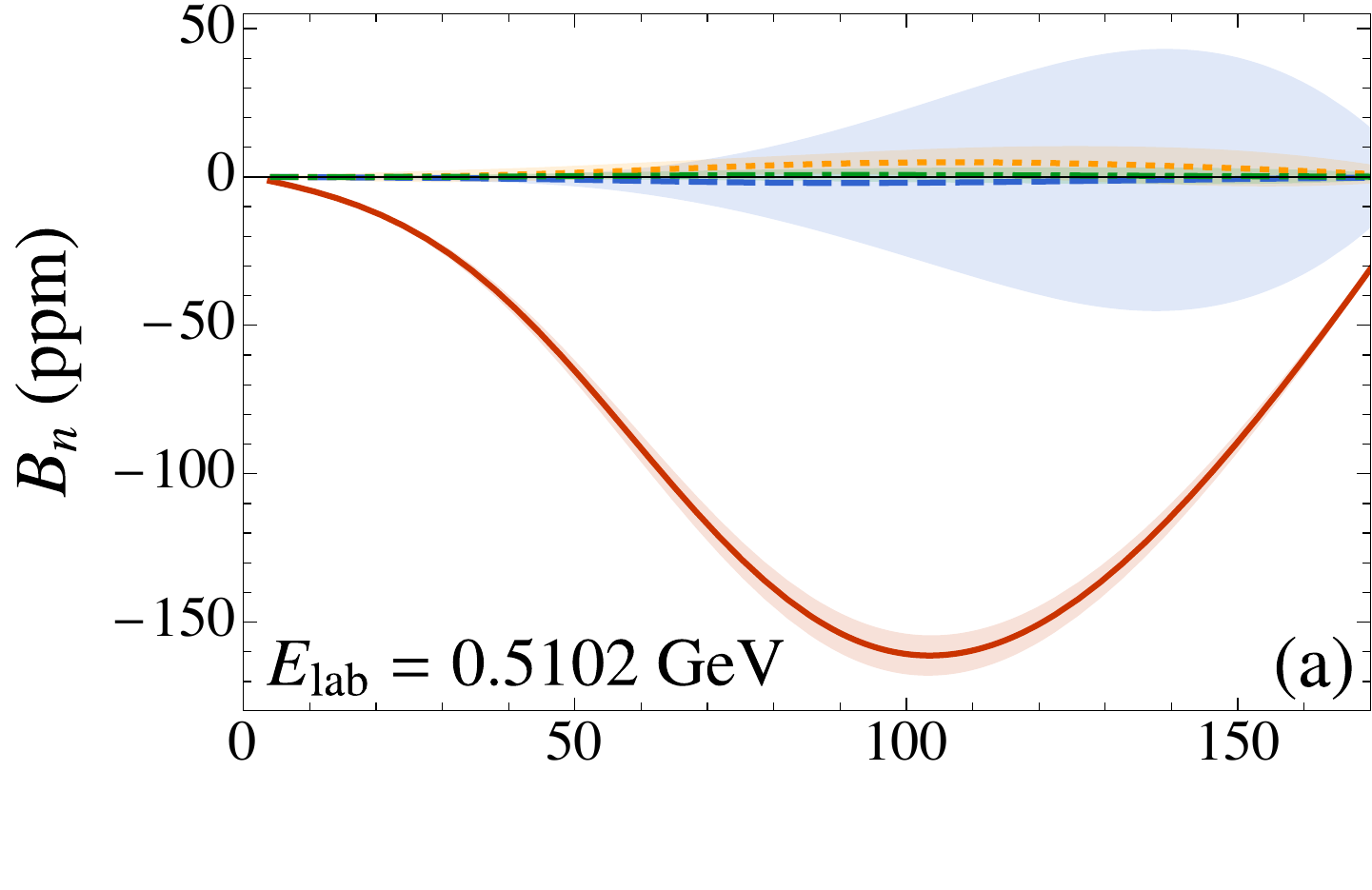}} \hfill
\subfloat{
\includegraphics[width=0.49\textwidth]{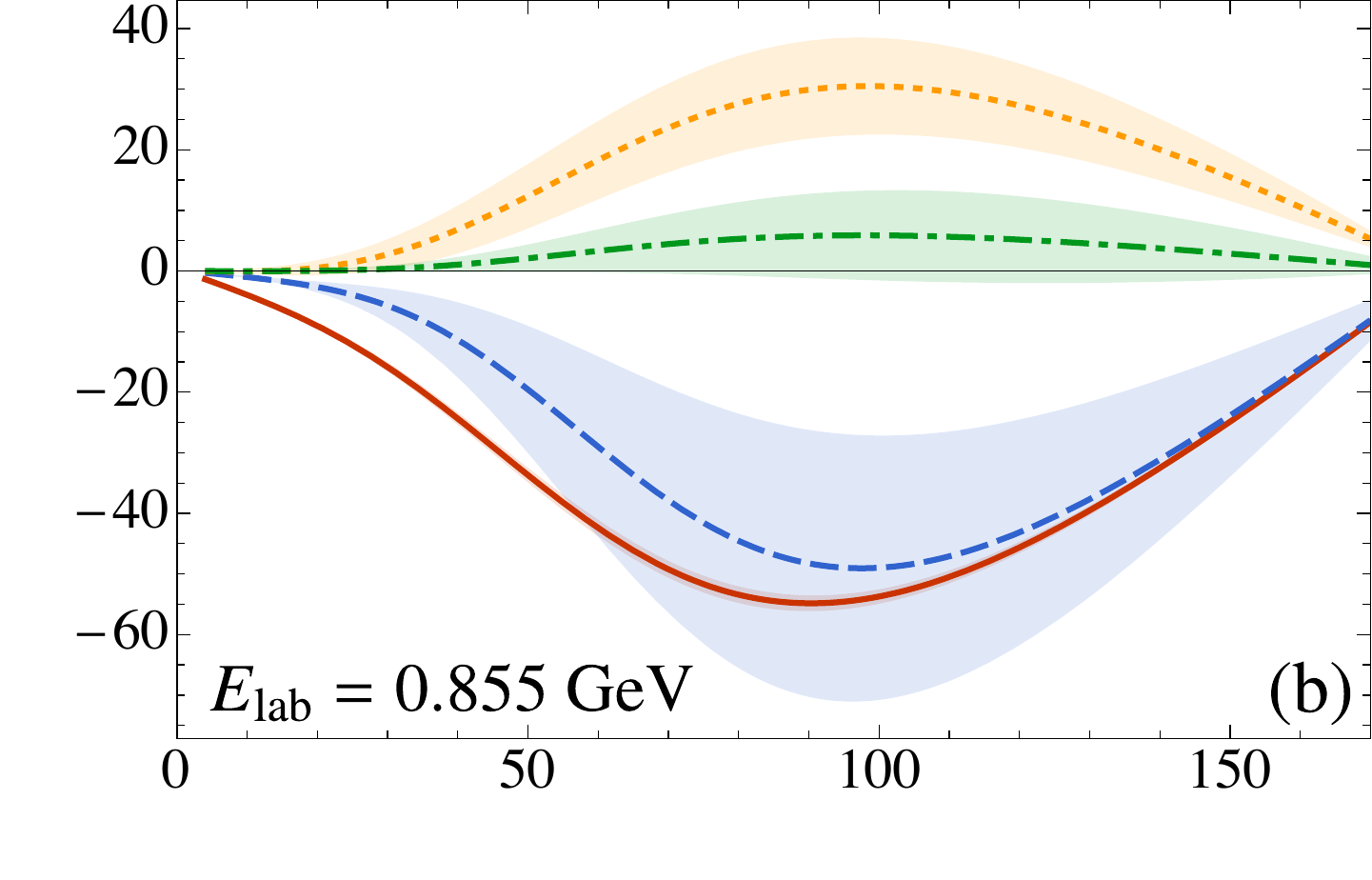}}\\
\subfloat{
\includegraphics[width=0.49\textwidth]{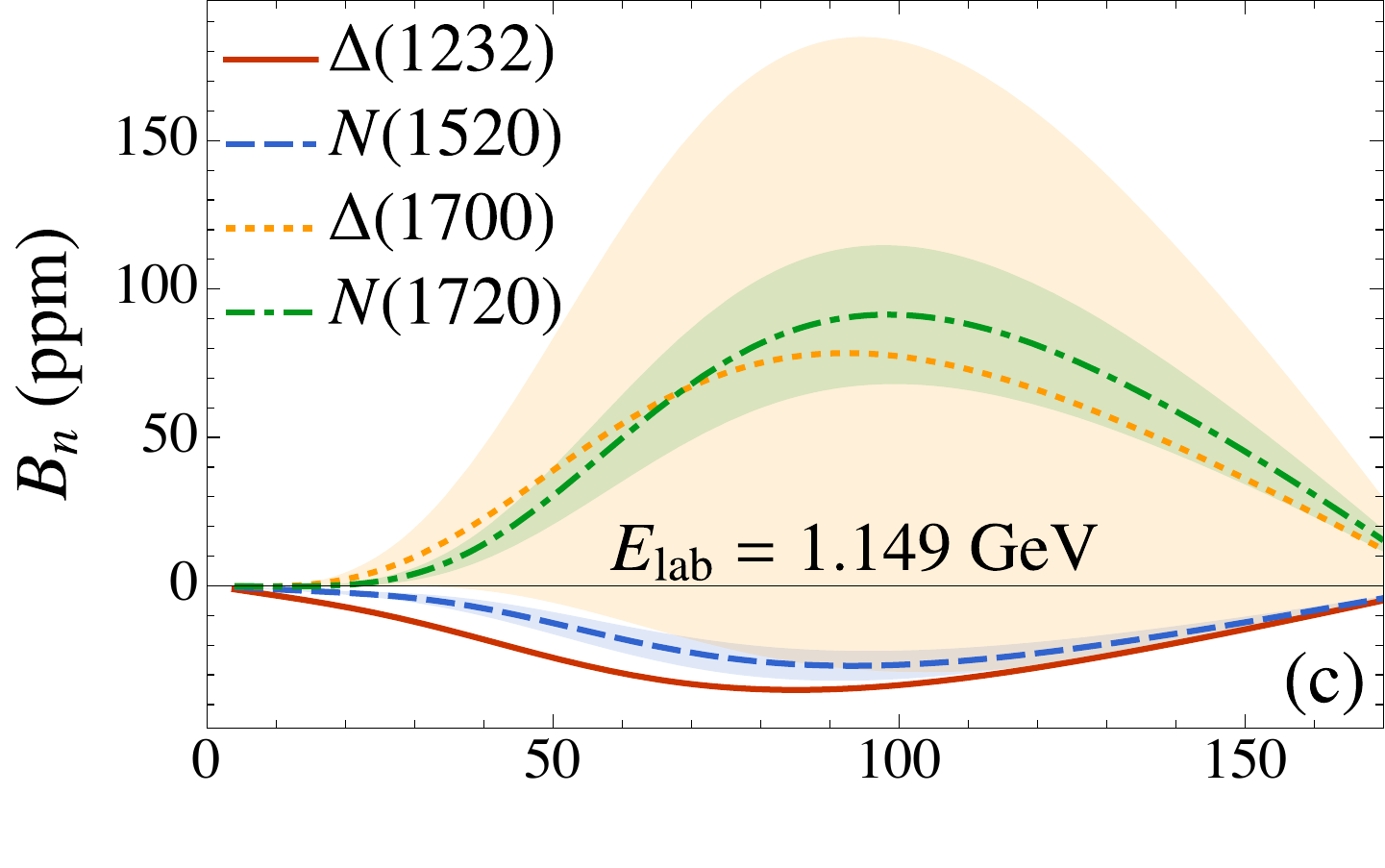}} \hfill
\subfloat{
\includegraphics[width=0.49\textwidth]{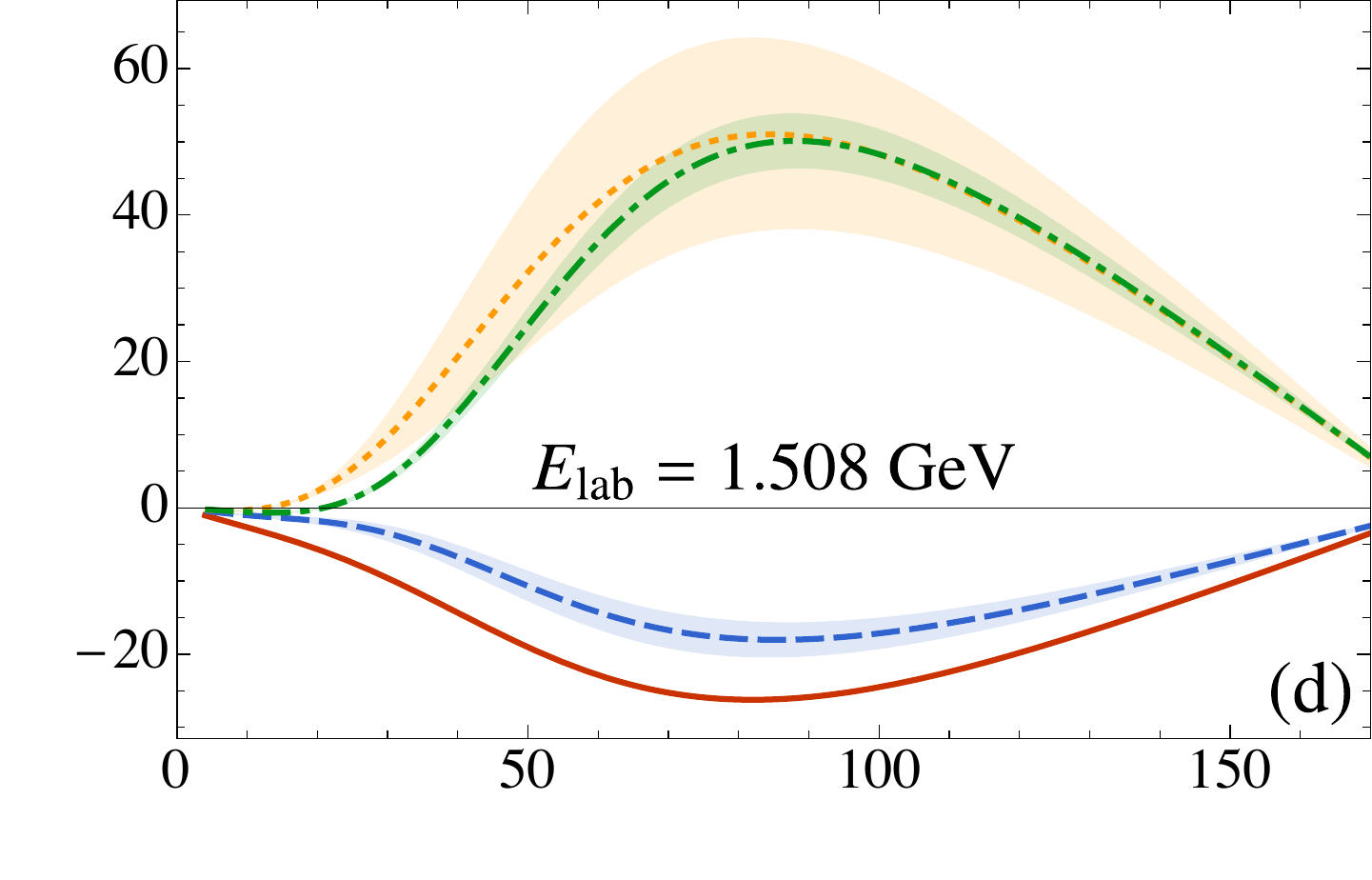}}\\
\includegraphics[width=0.49\textwidth]{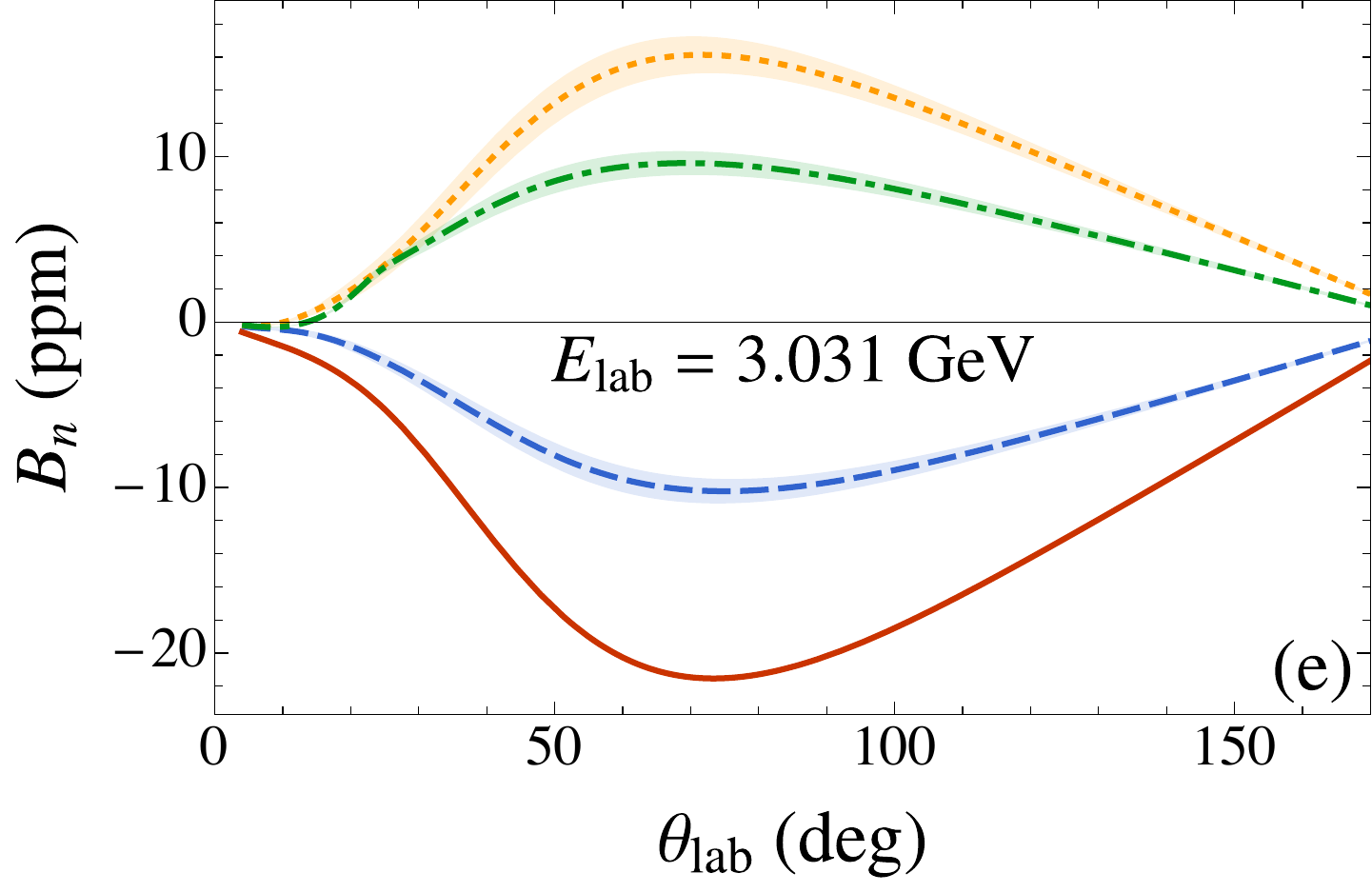} 
\caption{Resonance contributions to the beam normal SSA $B_n$ (in parts per million) as a function of scattering angle $\theta_{\textrm{lab}}$ at five representative beam energies $E_{\rm lab}$ equal to (a)~0.5102~GeV, (b)~0.855~GeV, (c)~1.149~GeV, (d)~1.508~GeV, and (e)~3.031~GeV. Only the four largest contributors are shown (from the $\Delta(1232)$, $N(1520)$, $\Delta(1700)$, and $N(1720)$), with the bands representing the uncertainty in the electrocouplings $A_h$.}
\label{fig.BnR}
\end{figure}

In this section we present the results for the beam normal SSA $B_n$, computed at beam energies relevant for existing experiments. 
To analyze the role of the resonances on the total SSA, in Fig.~\ref{fig.BnR} we illustrate the contributions to $B_n$ from the individual resonances at beam energies between $\approx 0.5$~GeV and $\approx 3$~GeV as function of the lab scattering angle $\theta_{\rm lab}$.

Among the resonances considered, the four spin-3/2 states $\Delta(1232)$, $N(1520)$, $\Delta(1700)$, and $N(1720)$ have sizeable effects, with some partial cancellation observed between them.
Contributions from resonances with spin 1/2 are smaller by at least an order of magnitude.
However, both the lower-mass spin-3/2 resonances $\Delta(1232)$ and $N(1520)$ give negative contributions to $B_n$, even though these states have different isospin and parity.
On the other hand, the two higher-mass spin-3/2 states $\Delta(1700)$ and $N(1720)$, with opposite parity and different isospin, make positive contributions to the total $B_n$.
No definite correlation between the isospin and parity is therefore observed in the imaginary part of the TPE amplitude for the case of normally polarized electrons elastically scattering from unpolarized protons.

At low beam energies, $E_{\rm lab} \lesssim 0.5$~GeV, the $\Delta(1232)$ state gives the dominant contribution to $B_n$ [Fig.~\ref{fig.BnR}(a)].
As the energy increases, the higher-mass resonances start playing a more significant role.
At $E_{\rm lab}= 0.855$~GeV, for example [Fig.~\ref{fig.BnR}(b)], the effect from the $N(1520)$, which has threshold energy $E_{\rm lab}^{\rm th} = 0.75$~GeV, becomes comparable to that of the $\Delta(1232)$.
It is interesting to note that the higher-mass resonance states $\Delta(1700)$ and $N(1720)$ show non-negligible effects even at beam energies below their excitation threshold (see Fig.~\ref{fig.BnR}(b)).
Such contributions, originating from the tail of the $W^2$ distribution for the nonzero width case, are not accounted for in the more approximate zero width calculations.
However, at energies above the threshold, the $\Delta(1700)$ and $N(1720)$ begin to dominate, as Figs.~\ref{fig.BnR}(c)-(e) demonstrate.
The dependence of these major resonances on the energy for fixed scattering angles will be further discussed below.
The overall magnitudes of the peak points of $B_n$ decrease with increasing beam energies for each of the resonances above their threshold, as evident from the scale of the panels in Fig.~\ref{fig.BnR}. 

It is also important to note that at forward laboratory scattering angles $\theta_{\rm lab}$, where most of the experimental data exist, the $\Delta(1232)$ contribution alone is a good approximation to the total, with the small effects from other resonances largely canceling in this region.
Furthermore, the elastic nucleon intermediate state gives a negligibly small effect in $B_n$, unlike the real part of the TPE amplitude in unpolarized $ep$ elastic scattering~\cite{Ahmed:2020uso}. 

\begin{figure}[t]%
\graphicspath{{ImagesK/}}
\centering
\subfloat{
\includegraphics[width=0.49\textwidth]{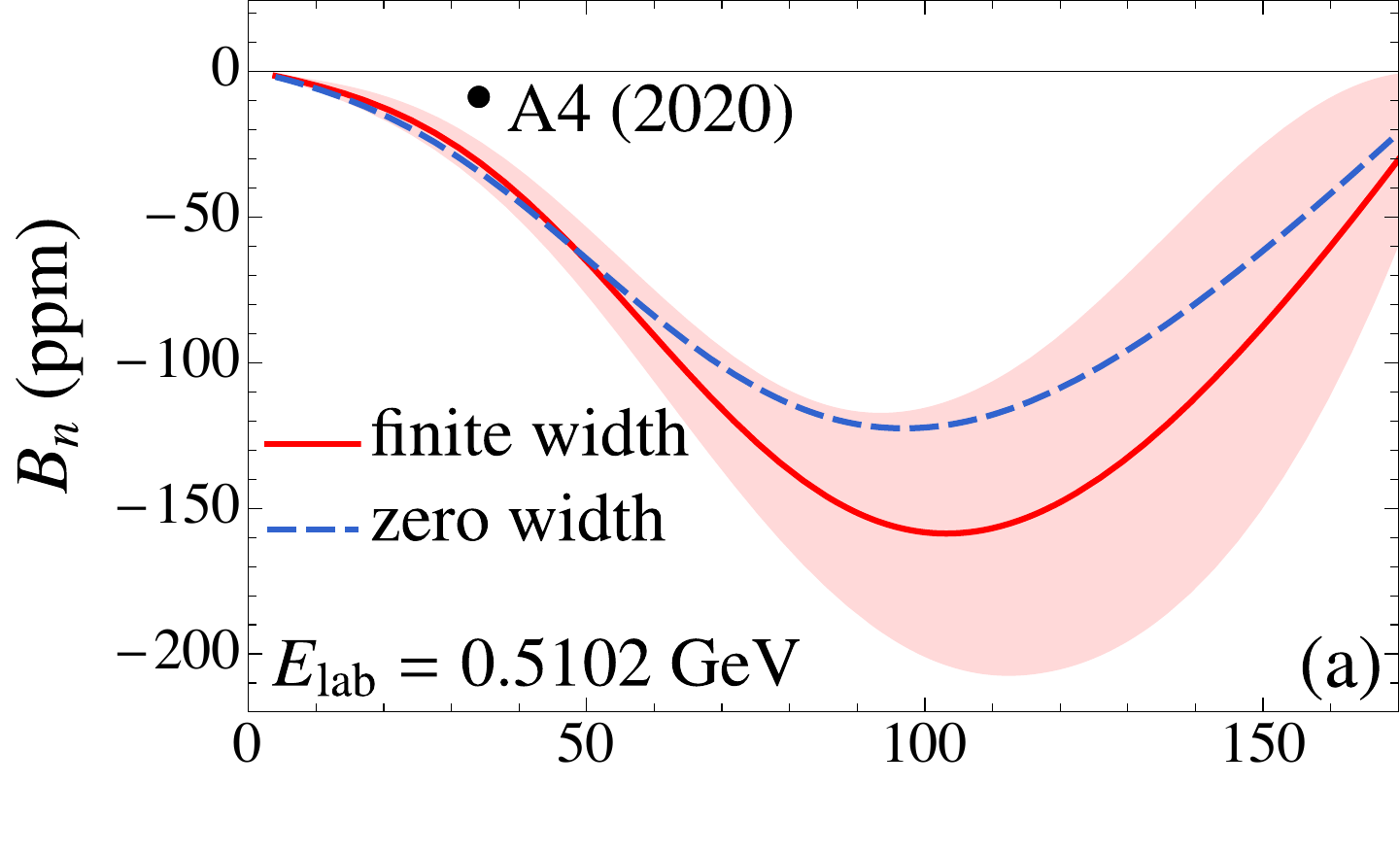}} \hfill
\subfloat{
\includegraphics[width=0.49\textwidth]{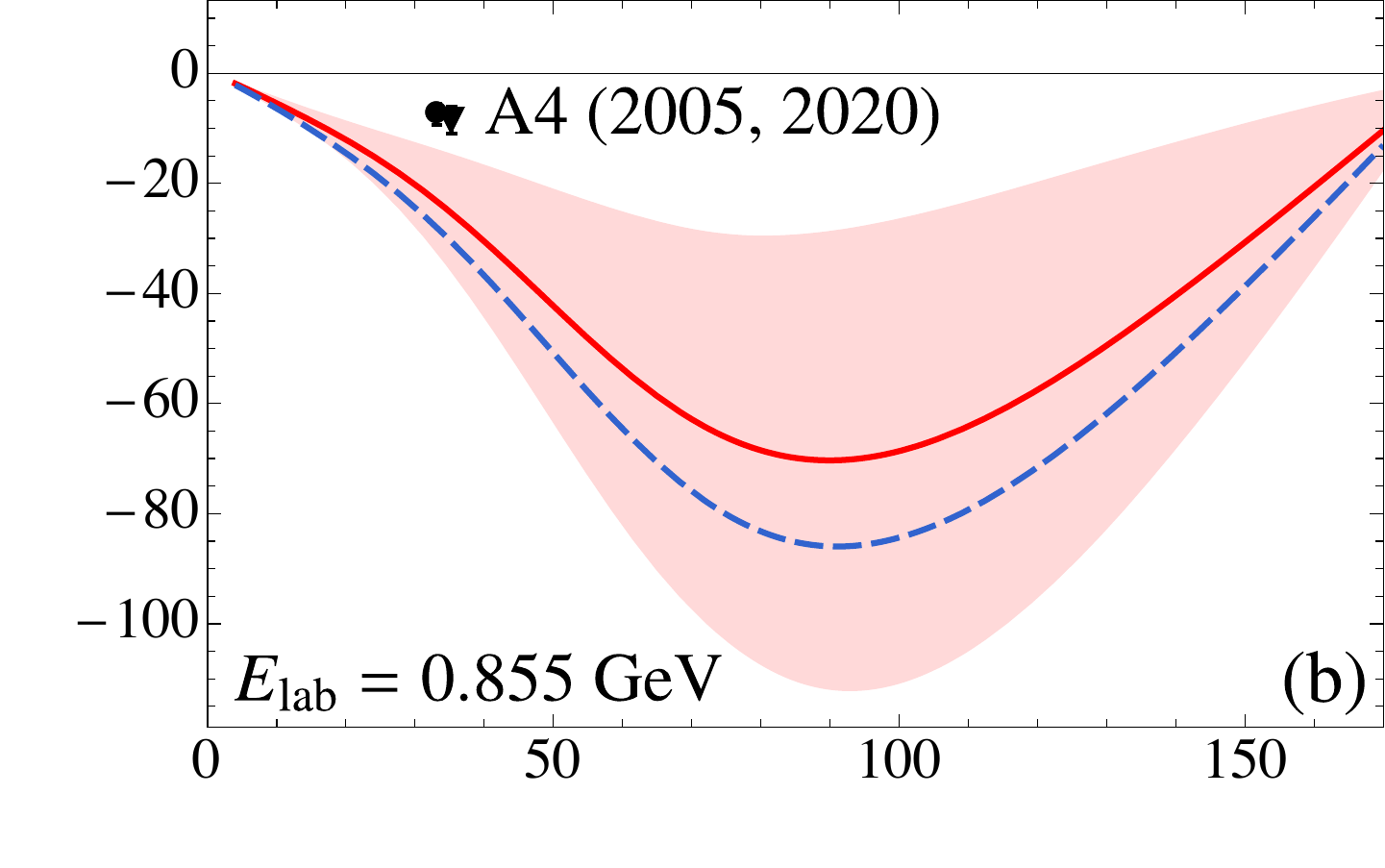}}\\
\subfloat{
\includegraphics[width=0.49\textwidth]{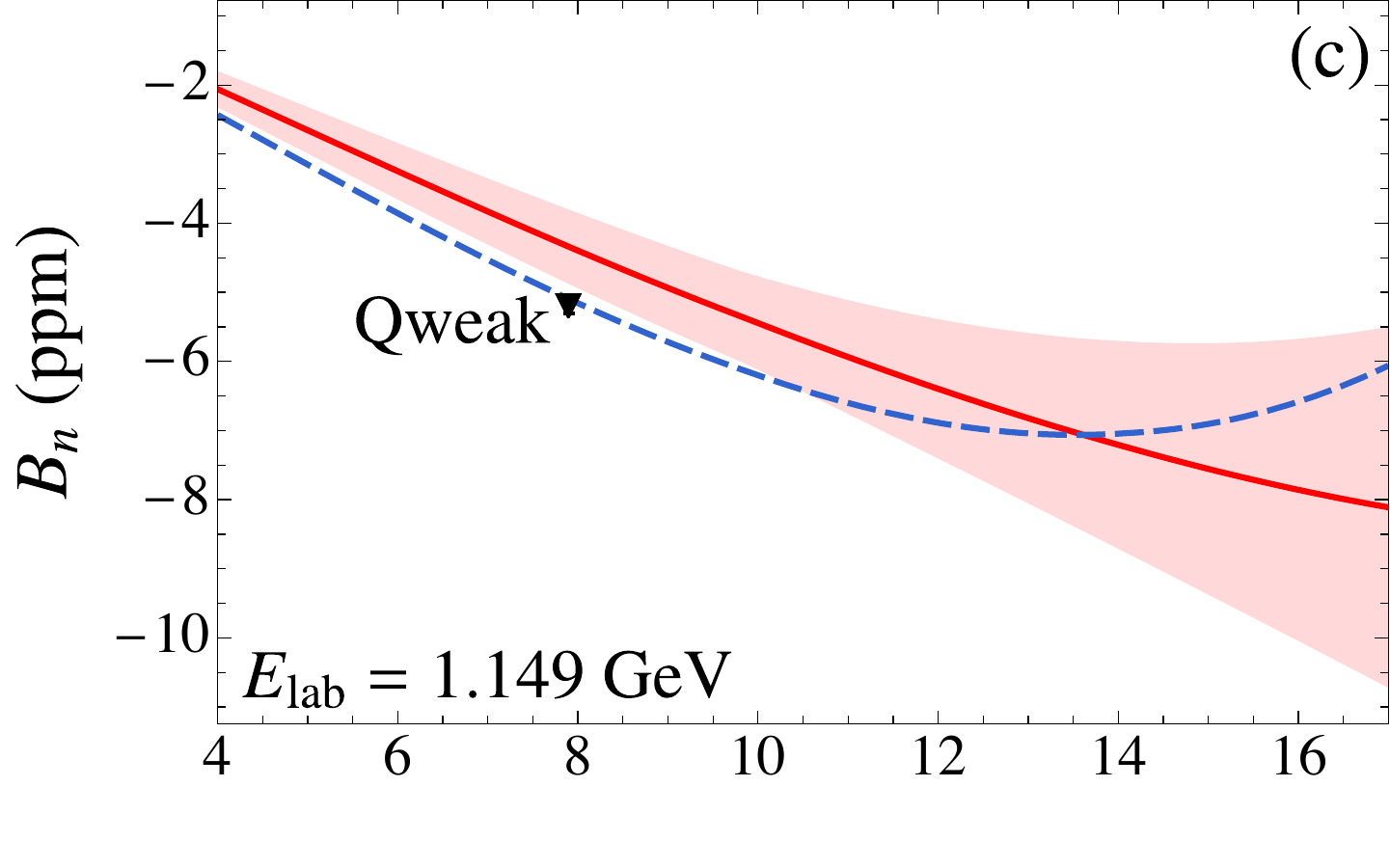}} \hfill
\subfloat{
\includegraphics[width=0.49\textwidth]{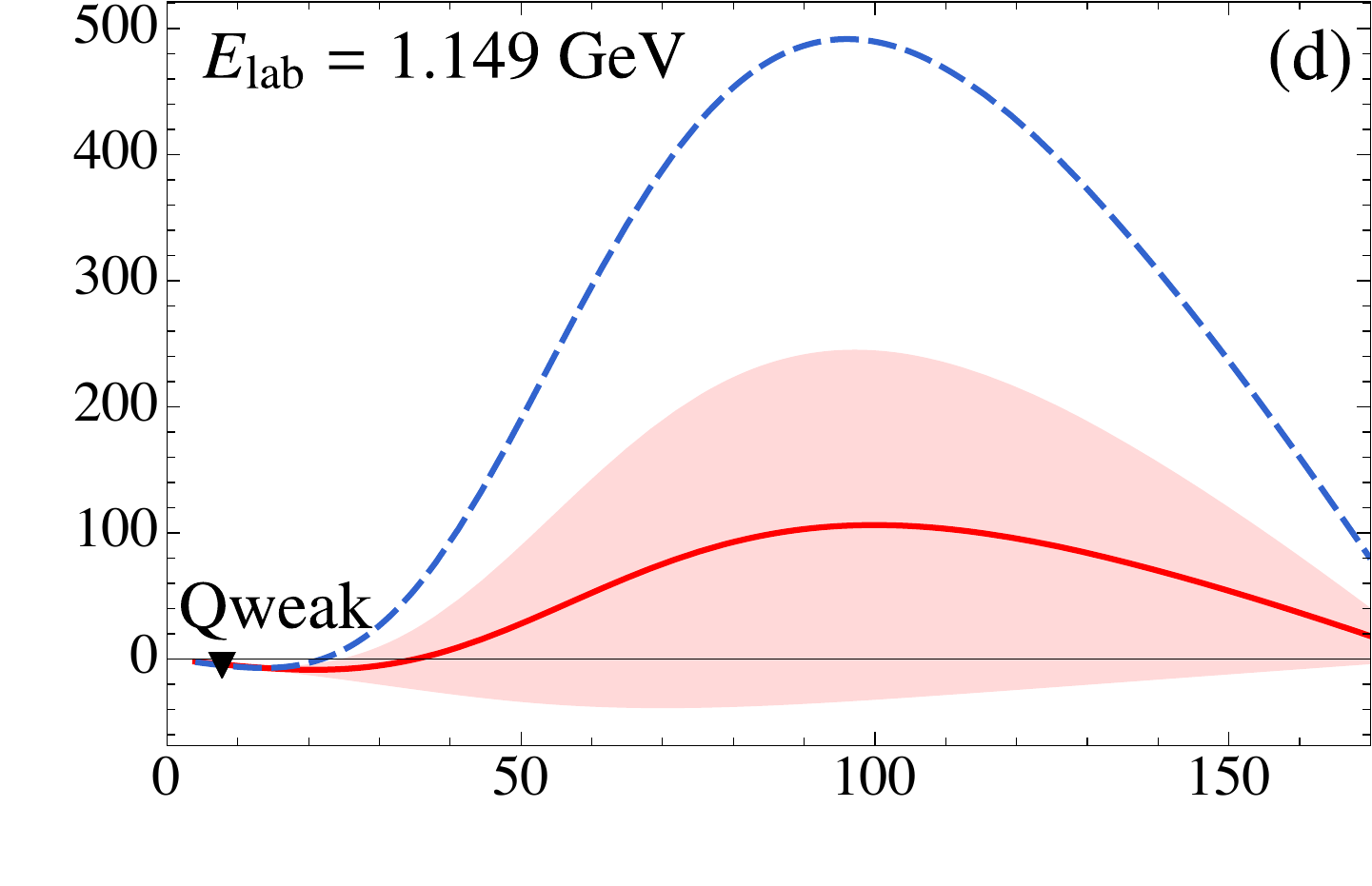}}\\
\subfloat{
\includegraphics[width=0.49\textwidth]{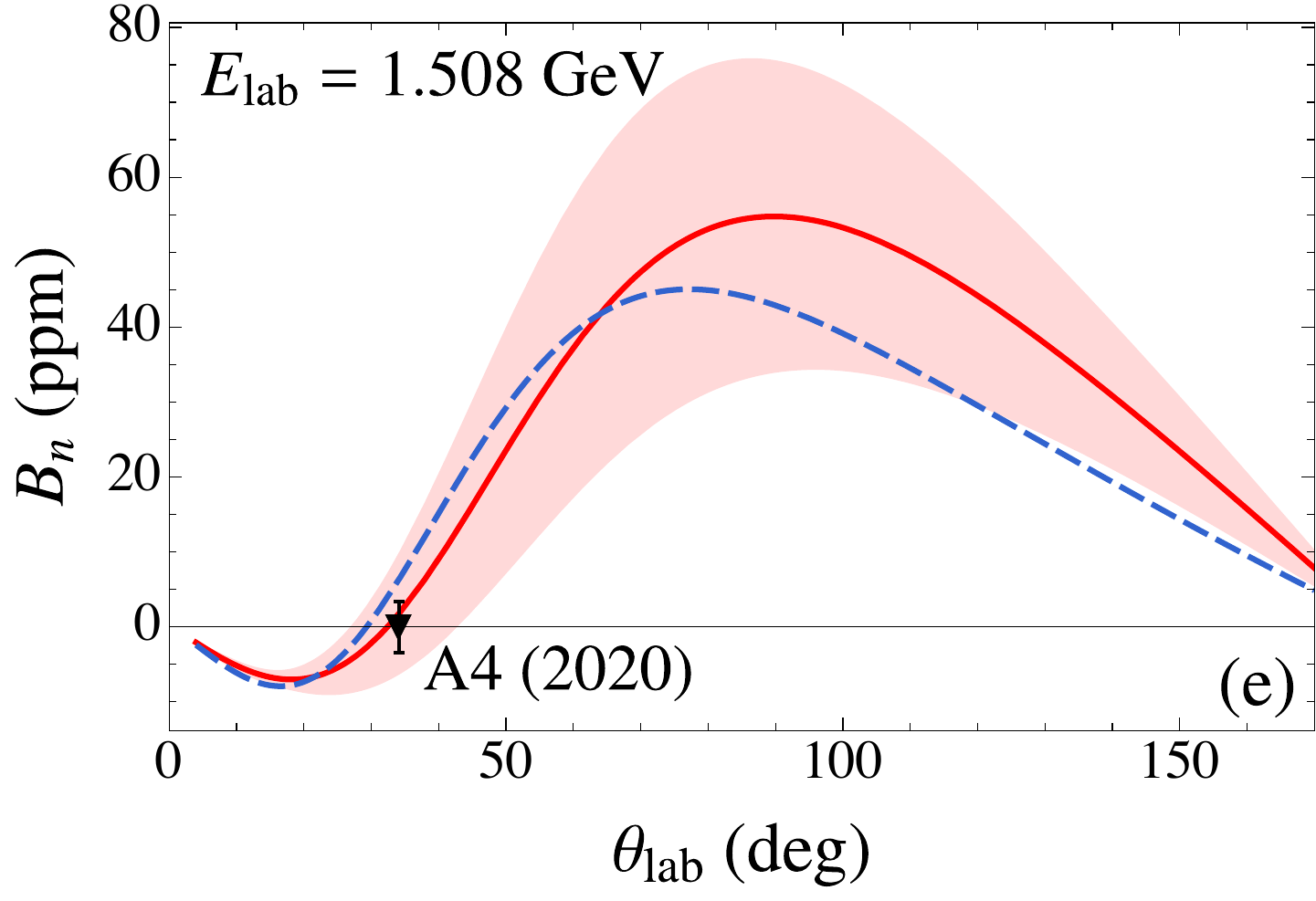}} \hfill
\subfloat{
\includegraphics[width=0.49\textwidth]{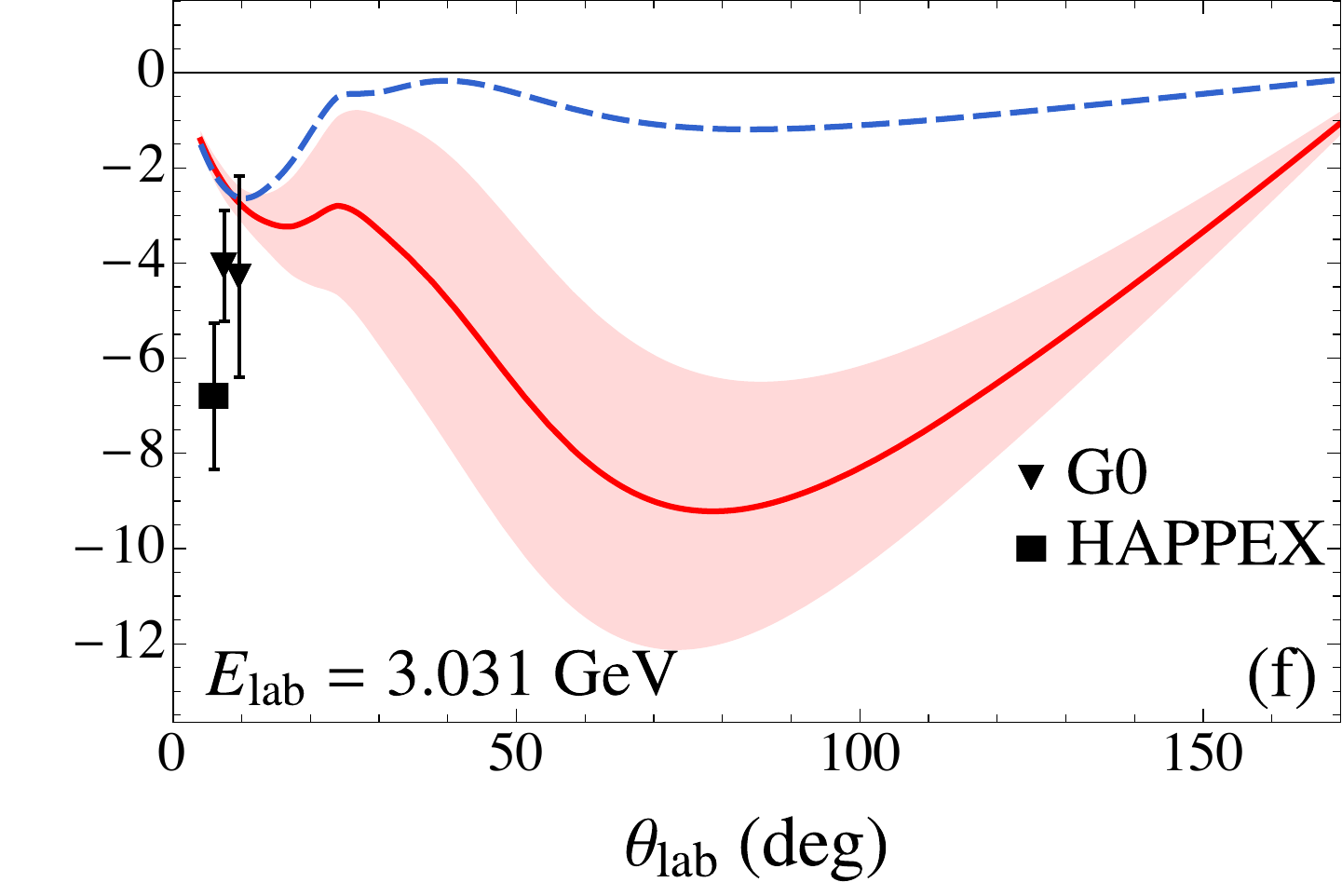}}
\caption{The total contribution (red solid lines) from the nucleon elastic and all nine resonance states in Table~\ref{tab.AhUncr} to the beam normal SSA $B_n$ as a function of $\theta_{\rm lab}$ for fixed beam energies corresponding to the A4 \cite{Gou:2020viq, Maas:2004pd}, Q$_{\rm weak}$ \cite{QWeak:2020fih}, G0 \cite{G0:2007mbp}, and HAPPEX \cite{HAPPEX:2012fud} experiments (black symbols). The results in the zero width approximation are shown for comparison (blue dashed lines).}
\label{fig.BnExp}
\end{figure}

The combined effect of all nine resonances listed in Table~\ref{tab.AhUncr}, along with the nucleon elastic contribution, on the total $B_n$ is illustrated in Fig.~\ref{fig.BnExp}, at the same kinematics as in Fig.~\ref{fig.BnR}.
The full results with the finite resonance decay widths are contrasted with the approximate results computed in the zero width approximation over the entire range of scattering angles $\theta_{\rm lab}$.
Overall, the finite width effect is small in the forward limit for all the considered beam energies, but the results of the two width approximations deviate in the far forward and backward angles.
We believe this may be attributable to a non-negligible contribution from the QRCS region with $W$ above or below the threshold value $W=M_R$, which is the only value of $W$ in the zero width case.

At the lower beam energies, $E_{\rm lab} = 0.5102$~GeV and 0.855~GeV, the overall $B_n$, including the effects of all elastic and resonance intermediate states, can be approximated by the $\Delta(1232)$ state alone. 
Over the entire range of scattering angles $\theta_{\rm lab}$ studied, the total $B_n$ remains negative, with peak magnitude of $\sim 160$~ppm and $\sim 70$~ppm for $E_{\rm lab} = 0.5102$~GeV and 0.855~GeV, respectively.
Compared with the experimental values, the calculated $B_n$ overshoots the asymmetries measured by the A4 Collaboration at MAMI at $\theta_{\rm lab} \approx 35^{\circ}$ \cite{Maas:2004pd, Gou:2020viq} [Fig.~\ref{fig.BnExp}(a), (b)].
On the other hand, the calculated $B_n$ is in good agreement with the high-precision Q$_{\rm weak}$ measurement \cite{QWeak:2020fih} at $E_{\rm lab} = 1.149$~GeV and $\theta_{\rm lab} = 7.9^\circ$, within uncertainties [Fig.~\ref{fig.BnExp}(c), (d)].
The effect of the finite width at the Q$_{\rm weak}$ energy is relatively small at forward angles [zoomed-in plot in Fig.~\ref{fig.BnExp}(c)], but results in a significantly reduced asymmetry at less forward angles, $\theta_{\rm lab} \gg 20^\circ-30^\circ$, compared with the zero width approximation.

Interestingly, the recent measurement of the asymmetry by the A4 Collaboration \cite{Gou:2020viq} at the larger beam energy $E_{\rm lab} = 1.508$~GeV and angle $\theta_{\rm lab} = 34.1^\circ$ shows excellent agreement with the calculation, especially for the finite width model.
As seen in Fig.~\ref{fig.BnExp}(d) and \ref{fig.BnExp}(e), the asymmetry changes sign to become positive at intermediate and backward scattering angles, $\theta_{\rm lab} \gtrsim 40^\circ$ in the $E_{\rm lab} \approx 1-1.5$~GeV range (see also Fig.~\ref{fig.BnE} below).
At beam energy $E_{\rm lab} \approx 3$~GeV, three data points are available from the G0 \cite{G0:2007mbp} and HAPPEX \cite{HAPPEX:2012fud} Collaborations in the forward angle region, $6^\circ \leq \theta_{\rm lab} \lesssim 10^\circ$. 
The calculated value of $B_n$ agrees with the sign of the measured asymmetry within the uncertainty range, but has slightly smaller magnitude for the HAPPEX data point in particular.
A complete list of experimental and calculated $B_n$ values is presented in Table~\ref{tab.BnThExp}, including also the early SAMPLE Collaboration result \cite{SAMPLE:2000hoe} at $E_{\rm lab} = 0.2$~GeV.

\begin{table}[t]
\renewcommand\arraystretch{.75}
\centering
\caption{Experimental and calculated beam normal SSA $B_n$ from various experiments, along with the corresponding kinematics in the lab frame. The calculated results include uncertainty estimates from the input helicity amplitudes, while the experimental results give both statistical and systematic uncertainties.}
\begin{tabular*}{\textwidth}{l @{\extracolsep{\fill}} lrlcc} 
\hline\hline
{~Experiment} & $E_\textrm{lab}$ & $\theta_\textrm{lab}$~  & ~~$Q^2$ &
\!Calculated $B_n$ & Experimental $B_n$    \\
& \!\!(GeV) & ($^\circ$)~ & (GeV$^2$) & (ppm) & (ppm) \\
%\toprule
\hline
{~SAMPLE (2001)~\cite{SAMPLE:2000hoe}} 
& 0.2 & 146.1  & ~~0.1    & $-40.5 \pm 4.5$  & $-15.4 \pm 5.4$ \\[4pt]
{~A4 (2005)~\cite{Maas:2004pd}} 
& 0.855 & 35.0 & ~~0.230  & $-25.1 \pm 10.1$ & $-8.52 \pm 2.31 \pm 0.87$ \\
& 0.569 & 35.0 & ~~0.106  & $-29.9 \pm 6.8$  & $-8.59 \pm 0.89 \pm 0.75$ \\[4pt]
{~G0 (2007)~\cite{G0:2007mbp}} 
& 3.031 & 7.5  & ~~0.15   & $-2.36 \pm 0.31$ & $-4.06 \pm 0.99 \pm 0.63$ \\ 
&  3.031 & 9.6 & ~~0.25   & $-2.73 \pm 0.35$ & $-4.28 \pm 1.87 \pm 0.98$ \\[4pt]
{~G0 (2011)~\cite{G0:2011chs}} 
& 0.362 & 108.0 & ~~0.22  & $-320 \pm 80$    & $-176.5 \pm 9.4$ \\ 
& 0.687 & 108.0 & ~~0.63  & $ -87 \pm 60$    & $-21 \pm 24$ \\[4pt]
{~HAPPEX (2012)~\cite{HAPPEX:2012fud}} 
& 3.026 & 6.0  & ~~0.099  & $-2.01 \pm 0.27$ & $-6.80 \pm 1.54$ \\[4pt]
{~A4 (2017)~\cite{Rios:2017vsw}} 
& 0.315 & 145.0 & ~~0.22  & $-201 \pm 88$    & $-94.83 \pm 6.02 \pm 4.07$ \\
& 0.420 & 145.0 & ~~0.350 & $-176 \pm 44$    & $-99.55 \pm 6.73 \pm 4.63$ \\[4pt]
{~A4 (2020)~\cite{Gou:2020viq}} 
& 0.315 & 34.1 & ~~0.032  & $-21 \pm 11$     & $-2.22 \pm 0.40 \pm 0.43$ \\ 
&  0.42 & 34.1 & ~~0.057  & $-34.5 \pm 9.4$  & $-6.88 \pm 0.53 \pm 0.42$ \\ 
& 0.510 & 34.1 & ~~0.082  & $-31.0 \pm 7.3$  & $-9.32 \pm 0.63 \pm 0.62$ \\ 
& 0.855 & 34.1 & ~~0.218  & $-24 \pm 10$     & $-7.46 \pm 1.22 \pm 1.55$ \\
& 1.508 & 34.1 & ~~0.613  & ~~~~$1.7 \pm 8.2 $   & $-0.06 \pm 2.89 \pm 1.90$ \\[4pt]
{~$Q_{\textrm{weak}}$ (2020)~\cite{QWeak:2020fih}} 
& 1.149 & 7.9 & ~~0.0248  & $-4.34 \pm 0.54$ & $-5.194 \pm 0.067 \pm 0.082$ \\[4pt]
\hline\hline
\end{tabular*}
\label{tab.BnThExp}
\end{table}

\begin{figure}[ht!]%
\graphicspath{{ImagesK/}}
\centering
\subfloat{
\includegraphics[width=0.49\textwidth]{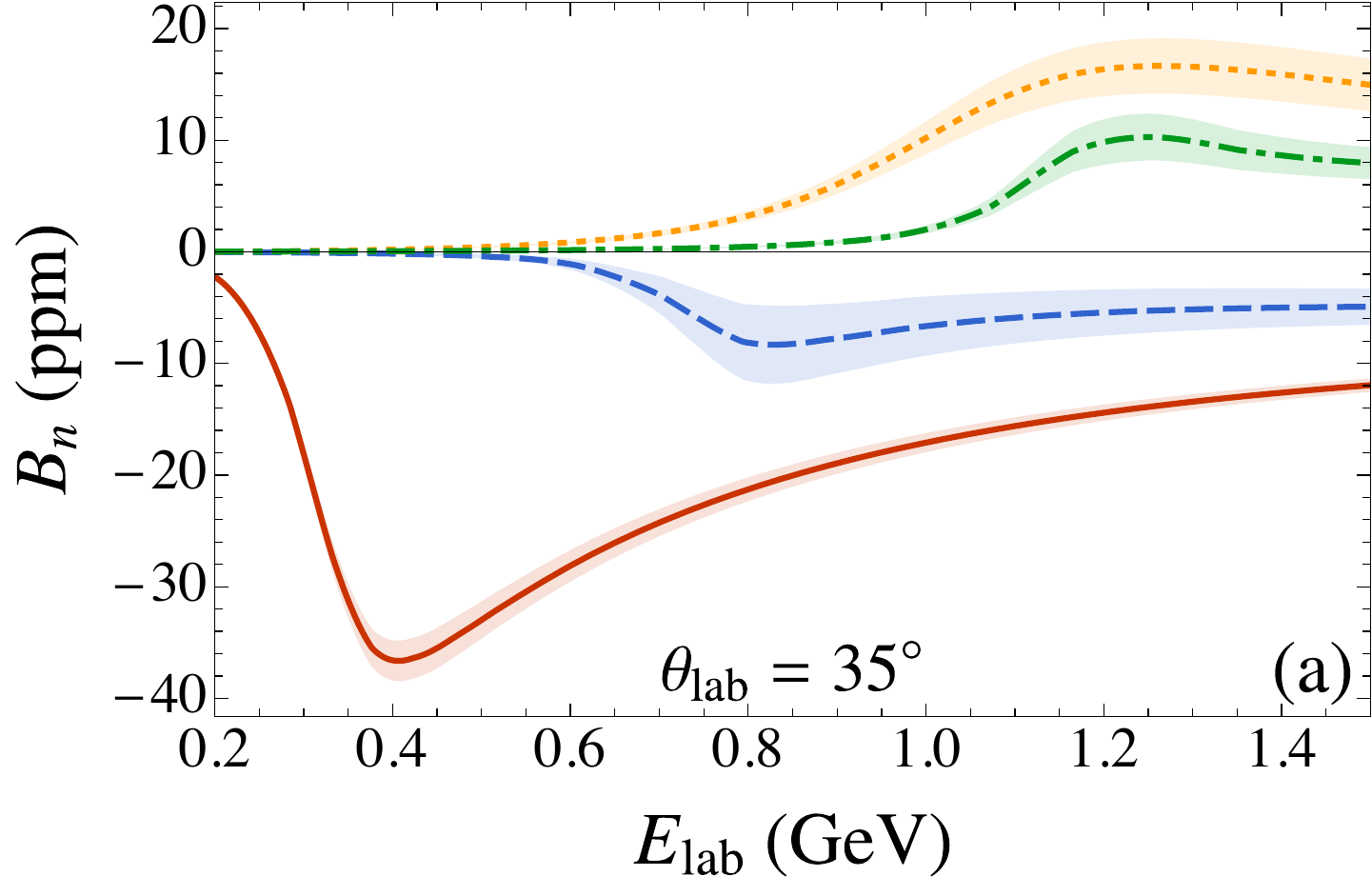}}
\hfill
\subfloat{
\includegraphics[width=0.49\textwidth]{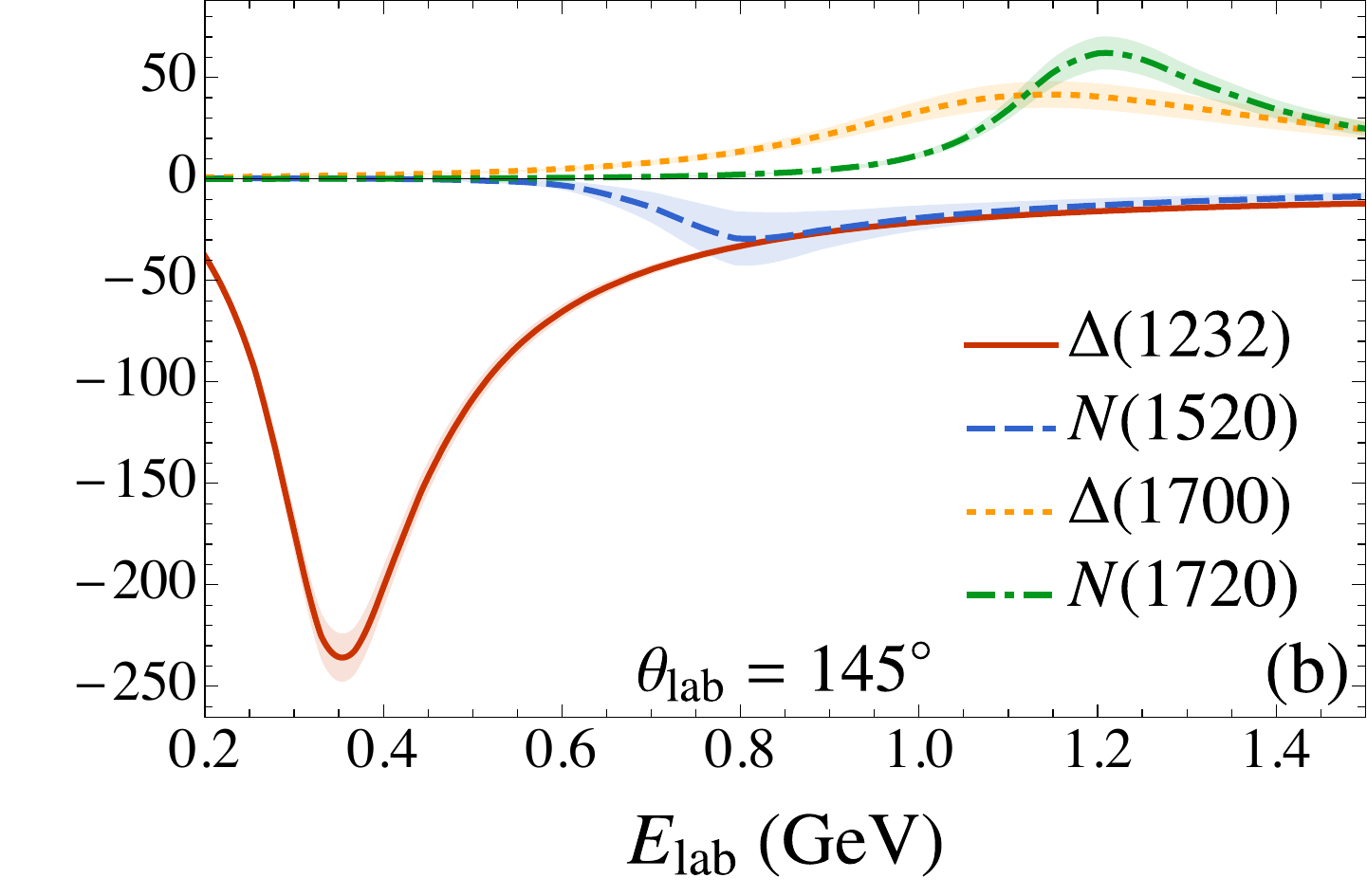}}\\
\subfloat{
\includegraphics[width=0.49\textwidth]{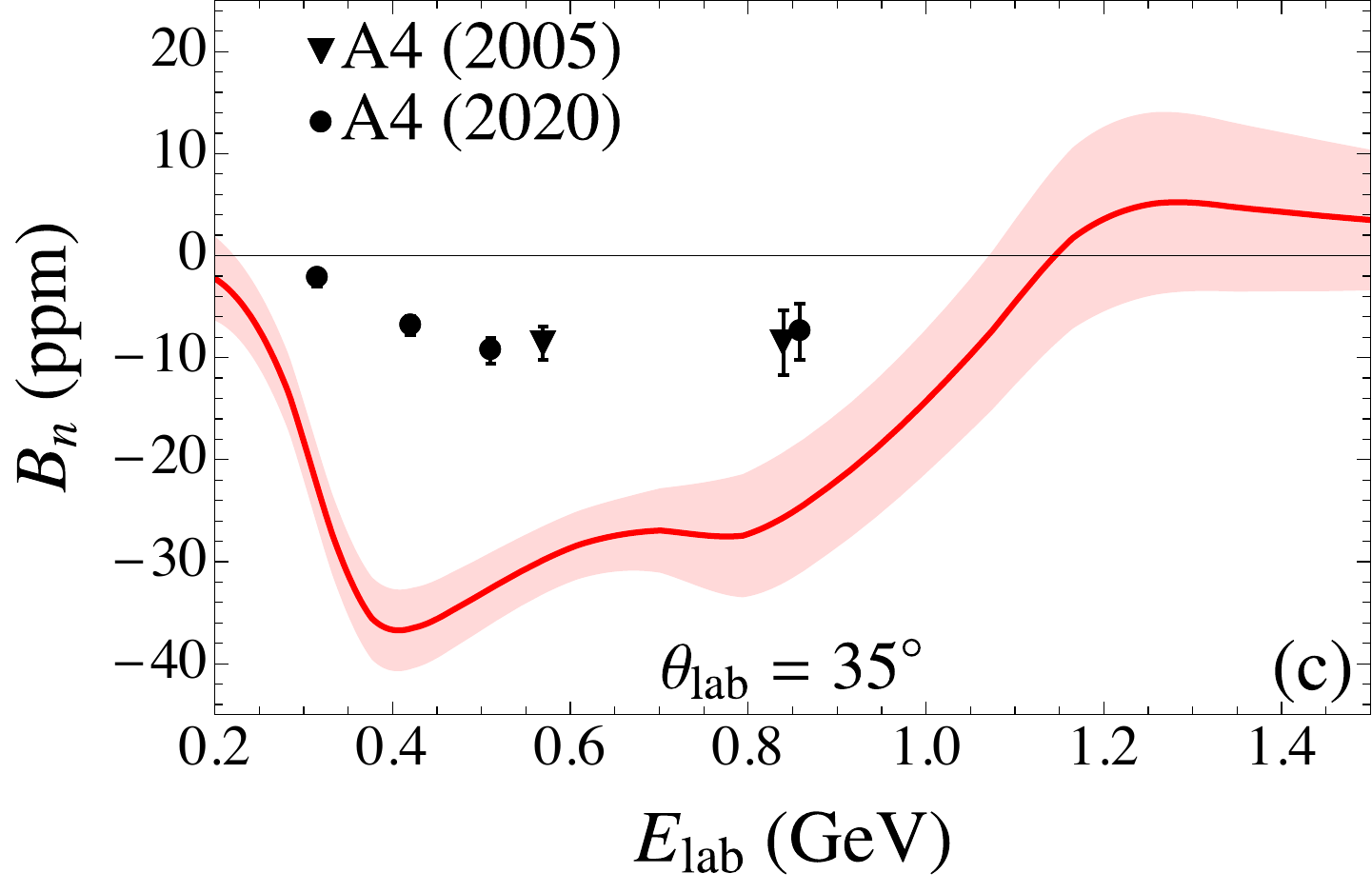}}
\hfill
\subfloat{
\includegraphics[width=0.49\textwidth]{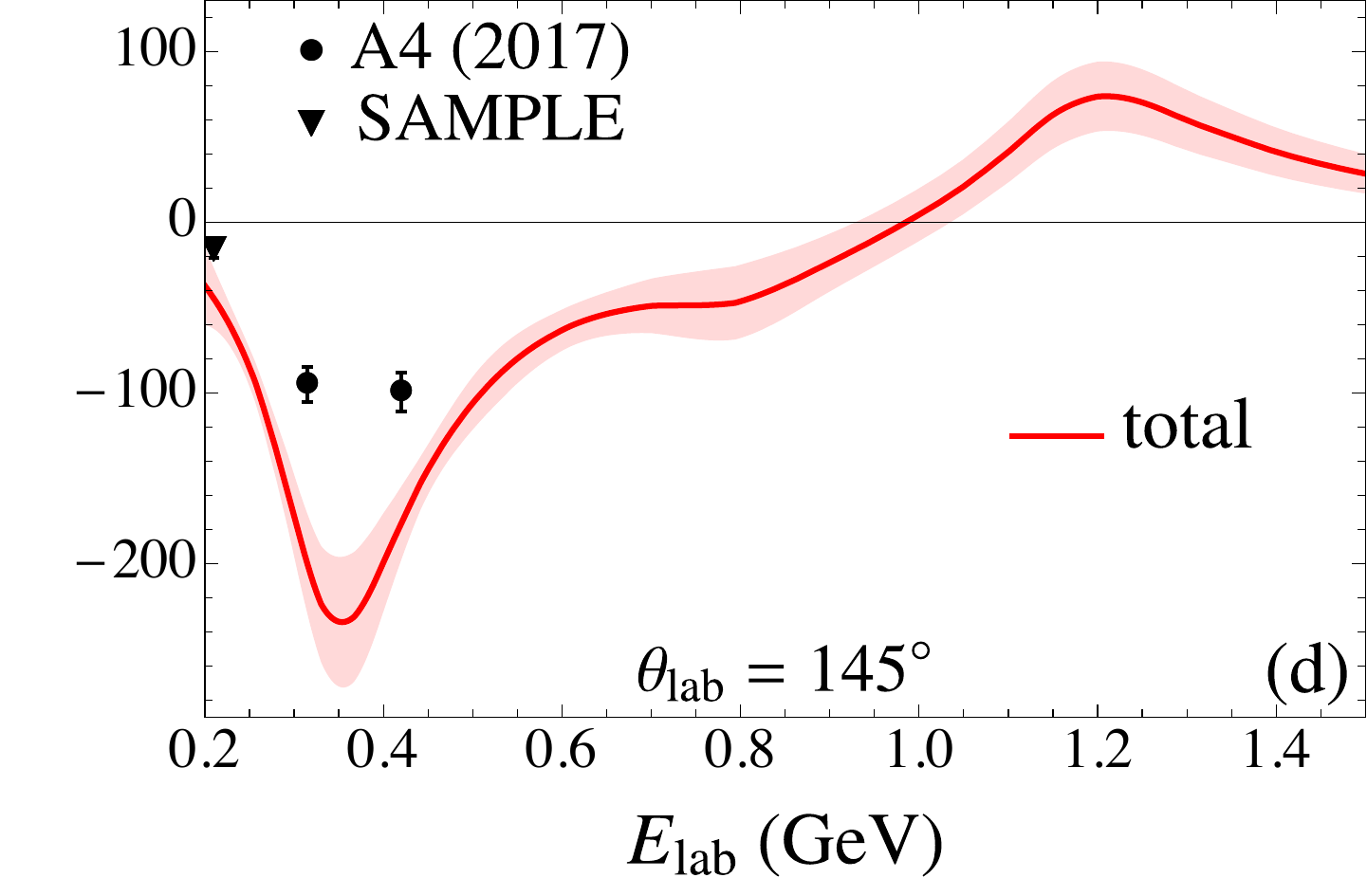}}
\caption{Beam normal SSA $B_n$ as a function of beam energy $E_{\rm lab}$ in the lab frame at representative scattering angles $\theta_{\rm lab} = 35^\circ$ and $145^\circ$. Contributions from the four largest contributors are shown in the top row while the bottom row represents the total $B_n$ from the nucleon plus all nine resonances. The experimental data points in the forward angle region are from A4 experiments~\cite{Maas:2004pd, Gou:2020viq}, and in the backward angle region from the SAMPLE~\cite{SAMPLE:2000hoe} and A4~\cite{Rios:2017vsw} experiments.}
\label{fig.BnE}
\end{figure}

To further illustrate the energy dependence of the total $B_n$ and its individual resonance contributions, we show in Fig.~\ref{fig.BnE} the asymmetry as a function of $E_{\rm lab}$ up to 1.5~GeV at the two representative scattering angles $\theta_{\rm lab} = 35^\circ$ and $145^\circ$ that are close to the experimental values.
The results illustrate again the dominance at low energies of the total asymmetry by the $\Delta(1232)$ state.
As expected, the higher mass resonances grow with increasing $E_{\rm lab}$, reaching their peak values at the threshold energies of the corresponding excited states, shown in Table~\ref{tab.AhUncr}.
After reaching the threshold limit, the positive contributions from the two heavier states $\Delta(1700)$ and $N(1720)$ outweigh the combined negative effects of the lower-mass states $\Delta(1232)$ and $N(1520)$, yielding a net positive value of $B_n$ at larger $E_{\rm lab}$. 
Compared with the experimental data from the SAMPLE experiment \cite{SAMPLE:2000hoe} and the series of measurements by the A4 Collaboration \cite{Maas:2004pd, Rios:2017vsw, Gou:2020viq}, the calculations give the same sign as the data in  Fig.~\ref{fig.BnE} in the measured region.
At the smaller scattering angle the calculation generally gives a larger magnitude for $B_n$ than that observed, while at the larger scattering angles the agreement between experiment and theory is reasonable, within uncertainties.  
The results suggest that, while the spin 1/2 and spin 3/2 resonances give  contributions to $B_n$ that have the correct sign and order of magnitude, there may still be room for higher spin states, such as spin 5/2 resonances, as well as nonresonant contributions to play some role.

% ......................................................................
\subsection{Target normal SSA $A_n$}
\label{ssec.AnResult}

\begin{figure}[t]
\graphicspath{{ImagesK/}}
\centering
\subfloat{
\includegraphics[width=0.49\textwidth]{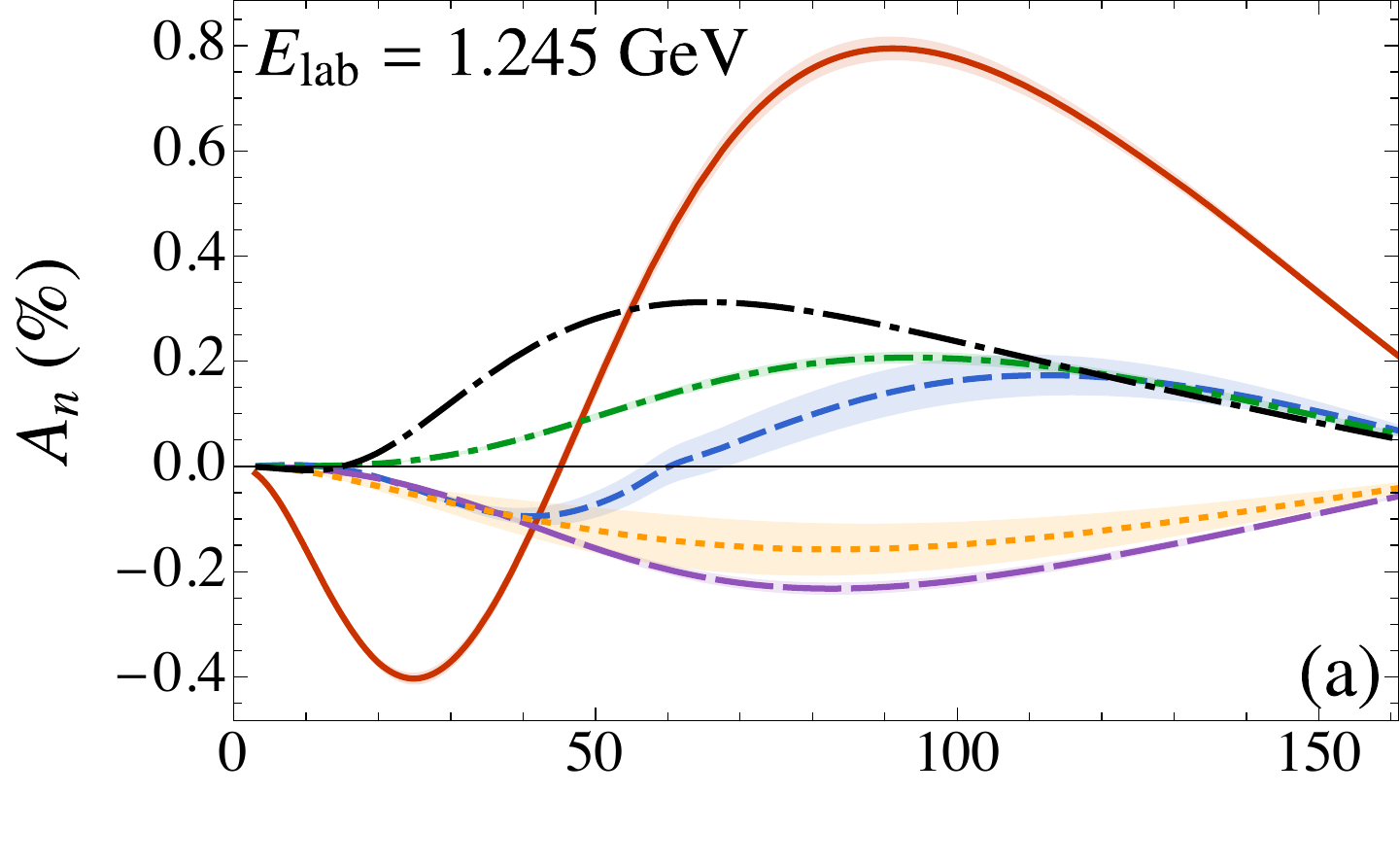}} \hfill
\subfloat{
\includegraphics[width=0.49\textwidth]{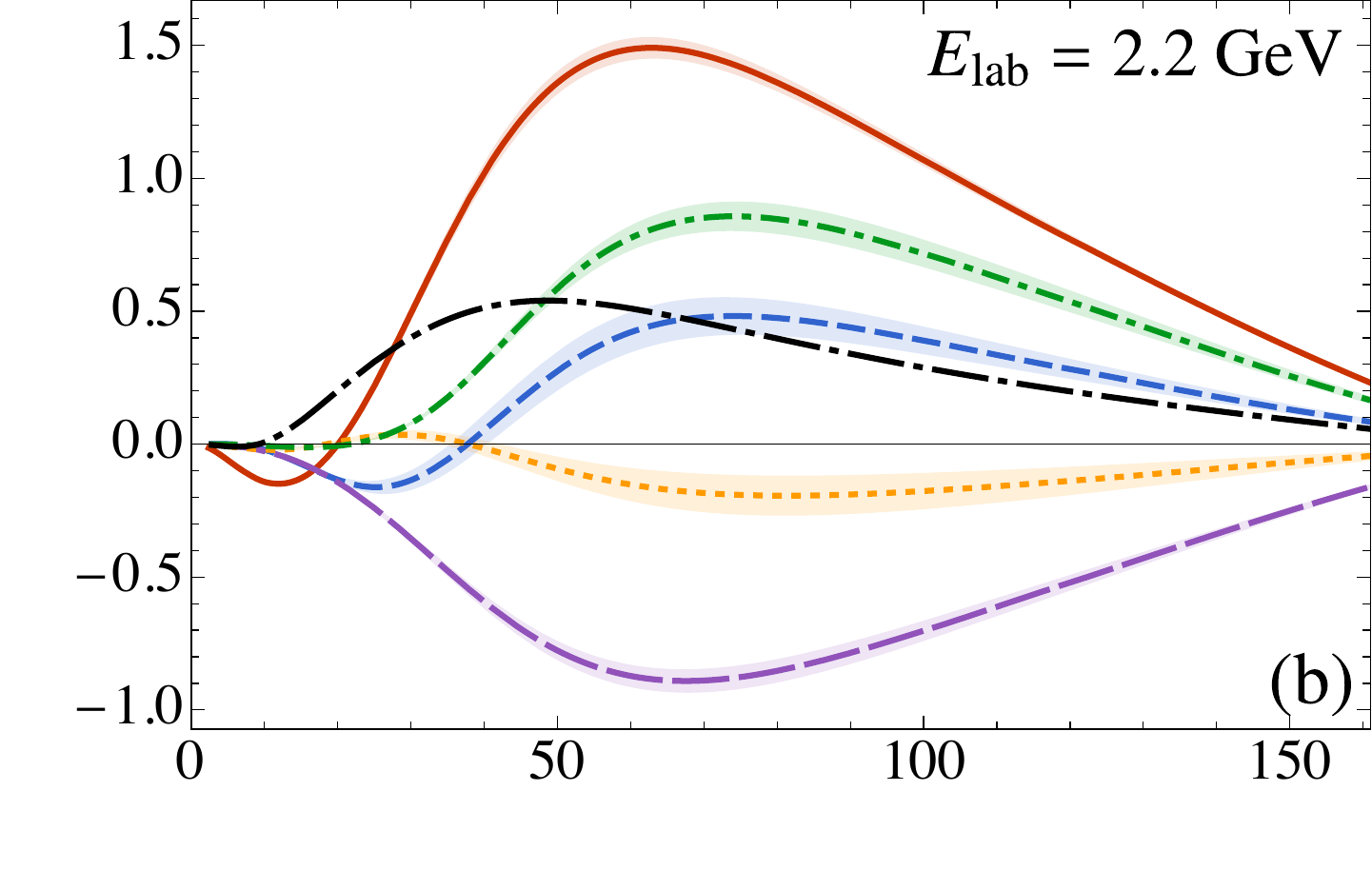}} \\
\subfloat{
\includegraphics[width=0.49\textwidth]{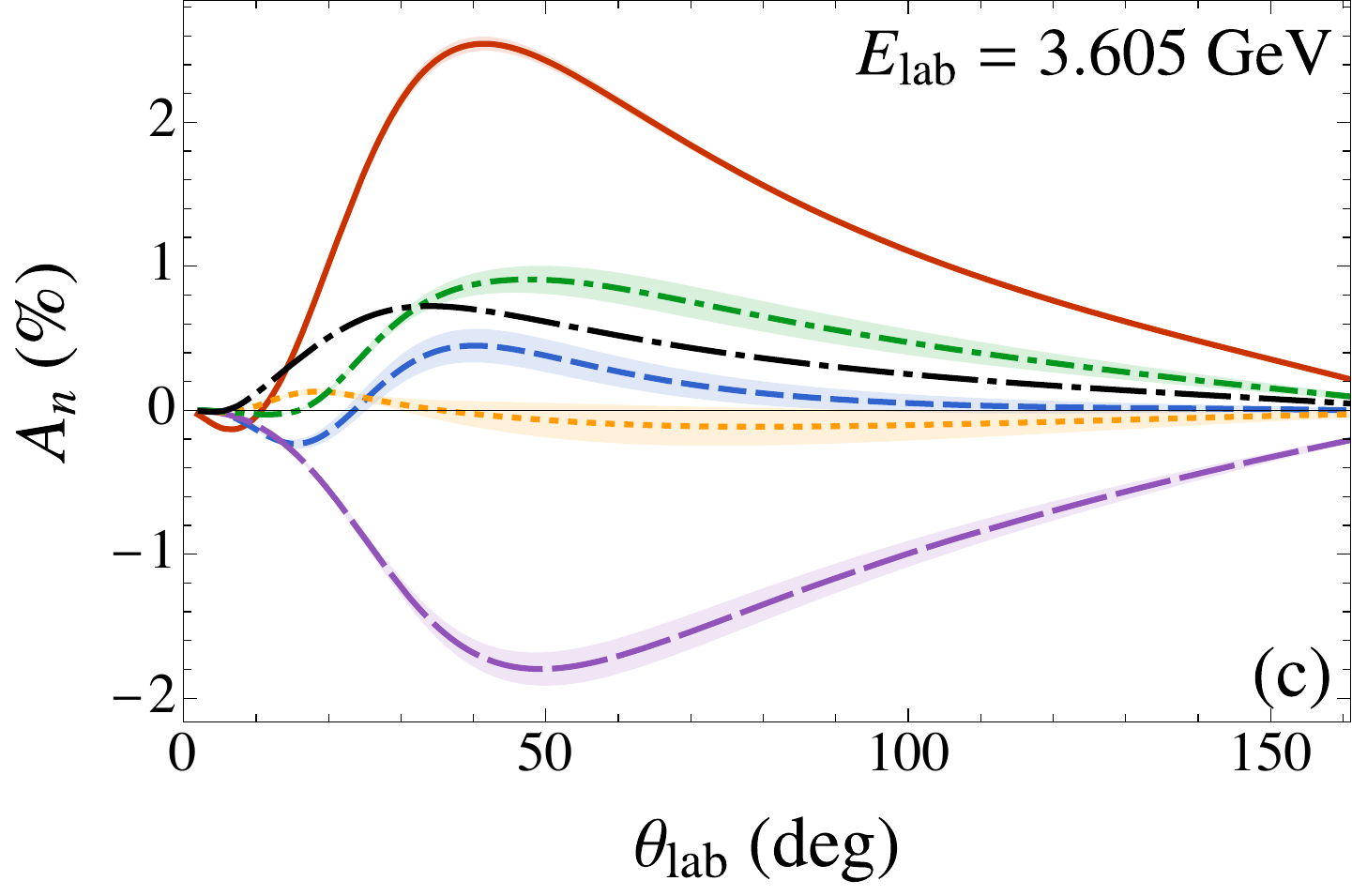}} \hfill
\subfloat{
\includegraphics[width=0.49\textwidth]{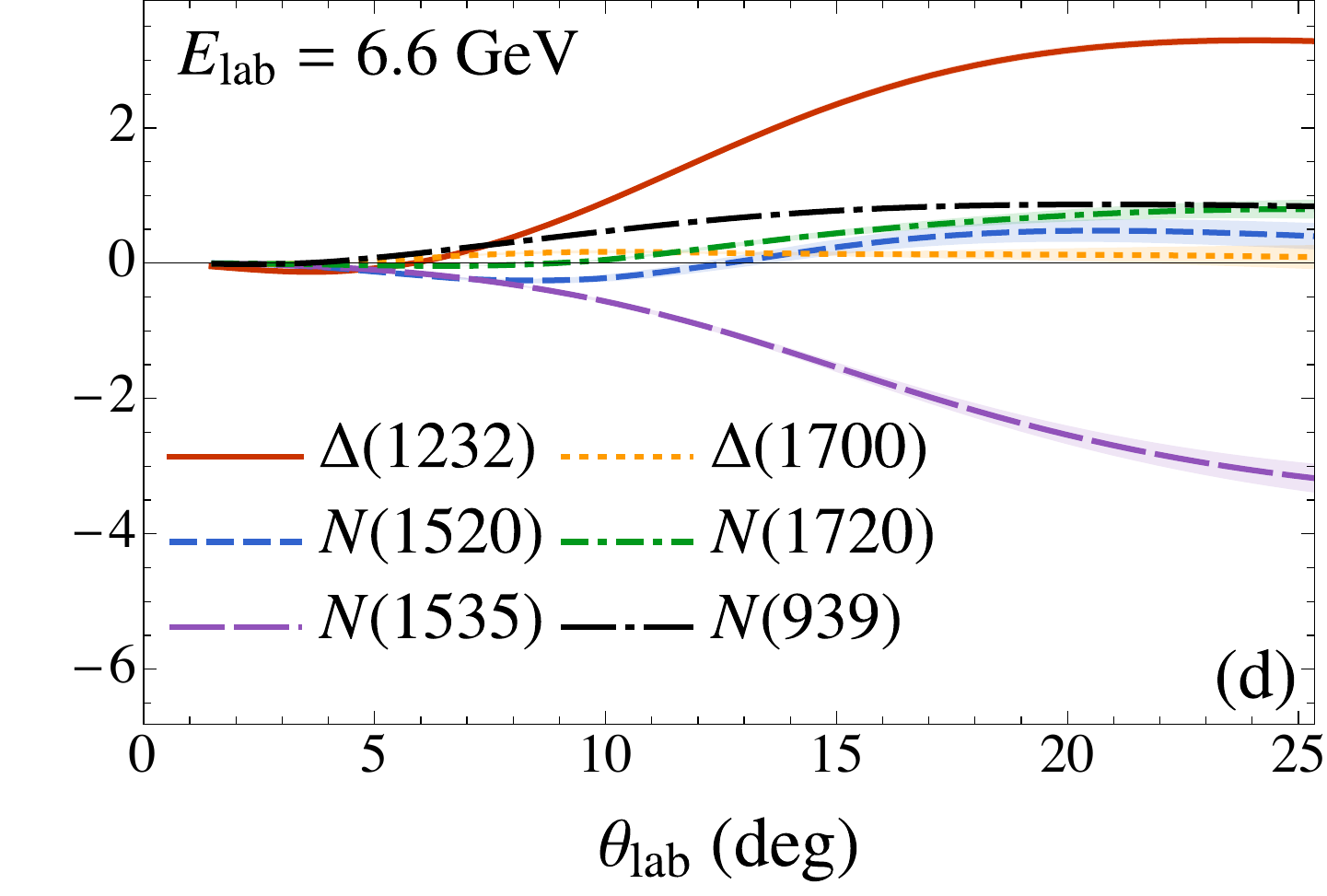}}
\caption{The five largest resonance contributions to the target normal SSA $A_n$ (in percent) as a function of the scattering angle $\theta_{\rm lab}$ at four representative beam energies $E_{\rm lab}$ given by (a)~1.245~GeV, (b)~2.2~GeV, (c)~3.605~GeV, and (d)~6.6~GeV.}
\label{fig.AnR}
\end{figure}

For the target normal SSA $A_n$, we consider four different beam energies, $E_{\rm lab} = 1.245$, 2.2, 3.605, and 6.6~GeV, corresponding to selected kinematics from the electron-$^3$He scattering experiment in Jefferson Lab Hall~A \cite{Zhang:2015kna, long2020}, and the proposed determination of the asymmetry in Ref.~\cite{Grauvogel:2021btg}. 
The contributions from the five major excited state resonances,
$\{ \Delta(1232)$, $N(1520)$, $N(1535)$, $\Delta(1700)$ and $N(1720) \}$,
to the total $A_n$ are shown in Fig.~\ref{fig.AnR} as function of the scattering angle $\theta_{\rm lab}$ at the chosen beam energies. 
For the highest energy $E_{\rm lab} = 6.6$~GeV, the asymmetry is shown up to a scattering angle $\theta_{\rm lab} \approx 25^\circ$, corresponding to $Q^2 = 5$~GeV$^2$, beyond which the hadronic approximation and the input electrocouplings parametrization used in the calculation are not expected to be reliable.

As anticipated, $A_n$ is in the sub-percent to percent range, and keeps increasing with beam energy in the far forward to backward directions, in contrast to the beam normal SSA $B_n$.
To further compare with $B_n$, we observe that the nucleon intermediate state alone has significant impact on the total $A_n$ for any value of $E_{\rm lab}$. 
Among the resonances, the $\Delta(1232)$ is again the dominant contributor over the entire range of $\theta_{\rm lab}$ and for all beam energies considered.
Particularly at forward angles, $\theta_{\rm lab} \lesssim 20^\circ$, the only sizeable contribution is that from the $\Delta(1232)$ state.
The effect of other resonances becomes comparable with the $\Delta(1232)$ at relatively larger scattering angles.

Interestingly, unlike for $B_n$ and the real part of the TPE correction~\cite{Ahmed:2020uso}, the contribution to the target normal SSA $A_n$ from the spin 3/2 nucleon state $N(1520)$ is relatively less significant for all beam energies $E_{\rm lab}$ considered over the entire range of $\theta_{\rm lab}$.
The two other spin 3/2 resonances, the $\Delta(1700)$ and $N(1720)$, have noticeable contributions at the lower beam energies, $E_{\rm lab} = 1.245$ and 2.2~GeV, but are of opposite sign, as shown in Fig.~\ref{fig.AnR}. 
At higher beam energies, the contribution to $A_n$ from these two states becomes negligible [Fig.~\ref{fig.AnR}(c), (d)]. 
On the other hand, the only spin 1/2 state, the $N(1535)$, is found to be a significant contributor to the total $A_n$.
As shown in Fig.~\ref{fig.AnR}(a), for $E_{\rm lab} = 1.245$~GeV the $A_n$ from the $N(1535)$ outweighs the contribution from all other states, with the exception of the $\Delta(1232)$.
With increasing $E_{\rm lab}$, the contribution from the $N(1535)$ rises even faster, almost negating the $\Delta(1232)$ contribution alone at the highest beam energy in Fig.~\ref{fig.AnR}(d). 
Considering all such partial cancellations, however, the sum of the elastic nucleon and $\Delta(1232)$ resonance contributions appears to be a good approximation to the total $A_n$.

\begin{figure}[ht!]%
\graphicspath{{ImagesK/}}
\centering
\subfloat{
\includegraphics[width=0.49\textwidth]{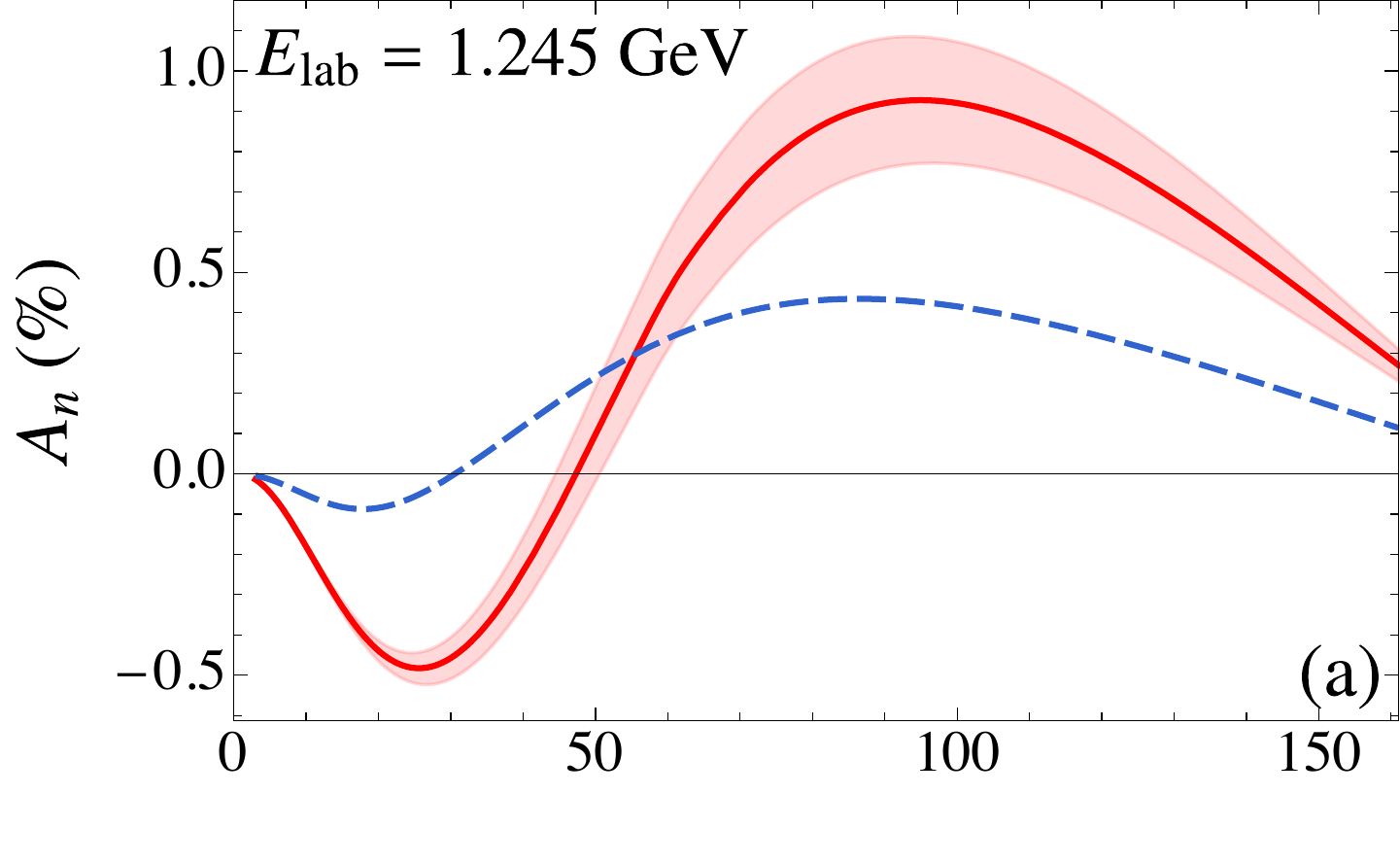}} \hfill
\subfloat{
\includegraphics[width=0.49\textwidth]{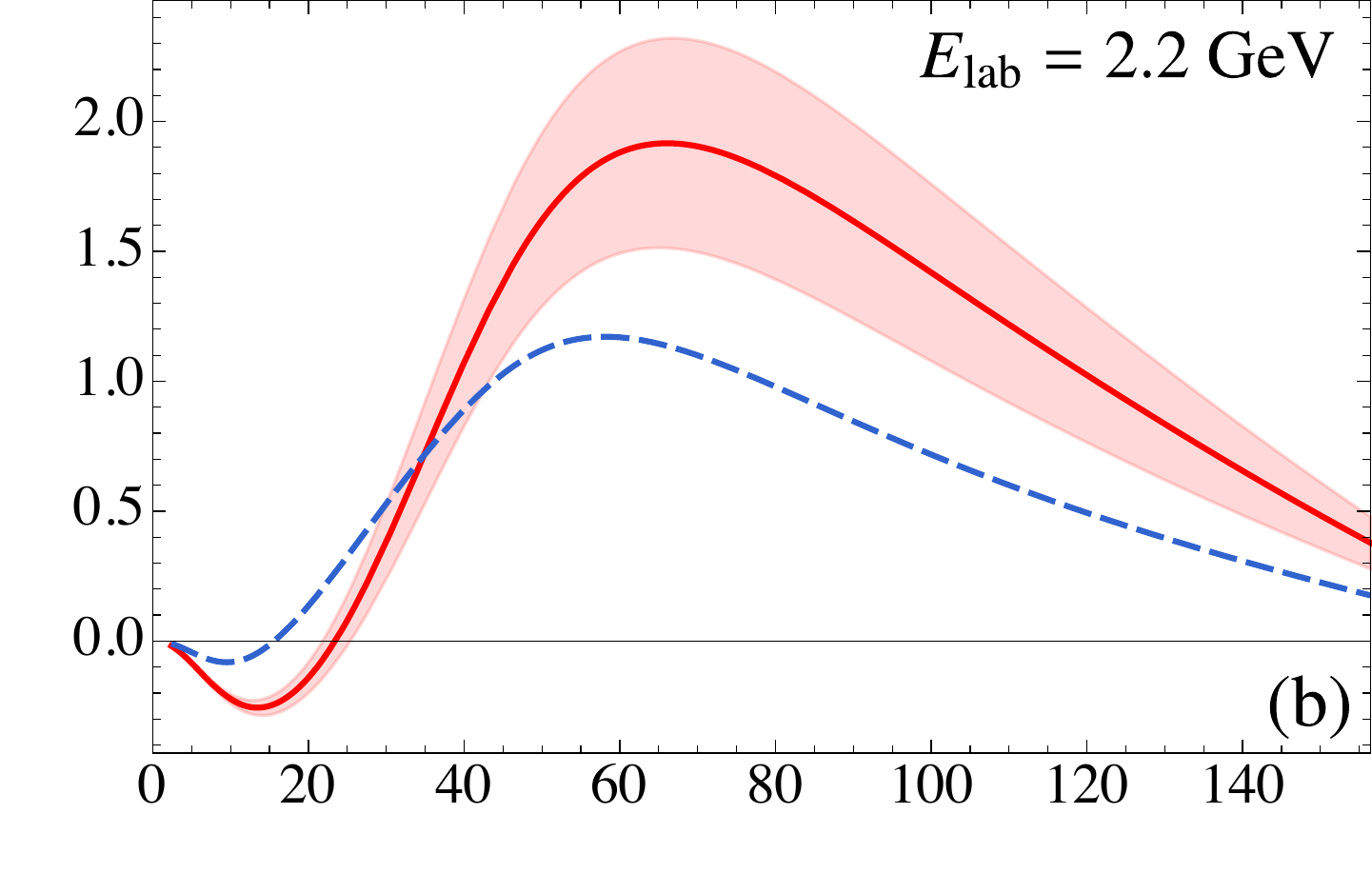}} \\
\subfloat{
\includegraphics[width=0.49\textwidth]{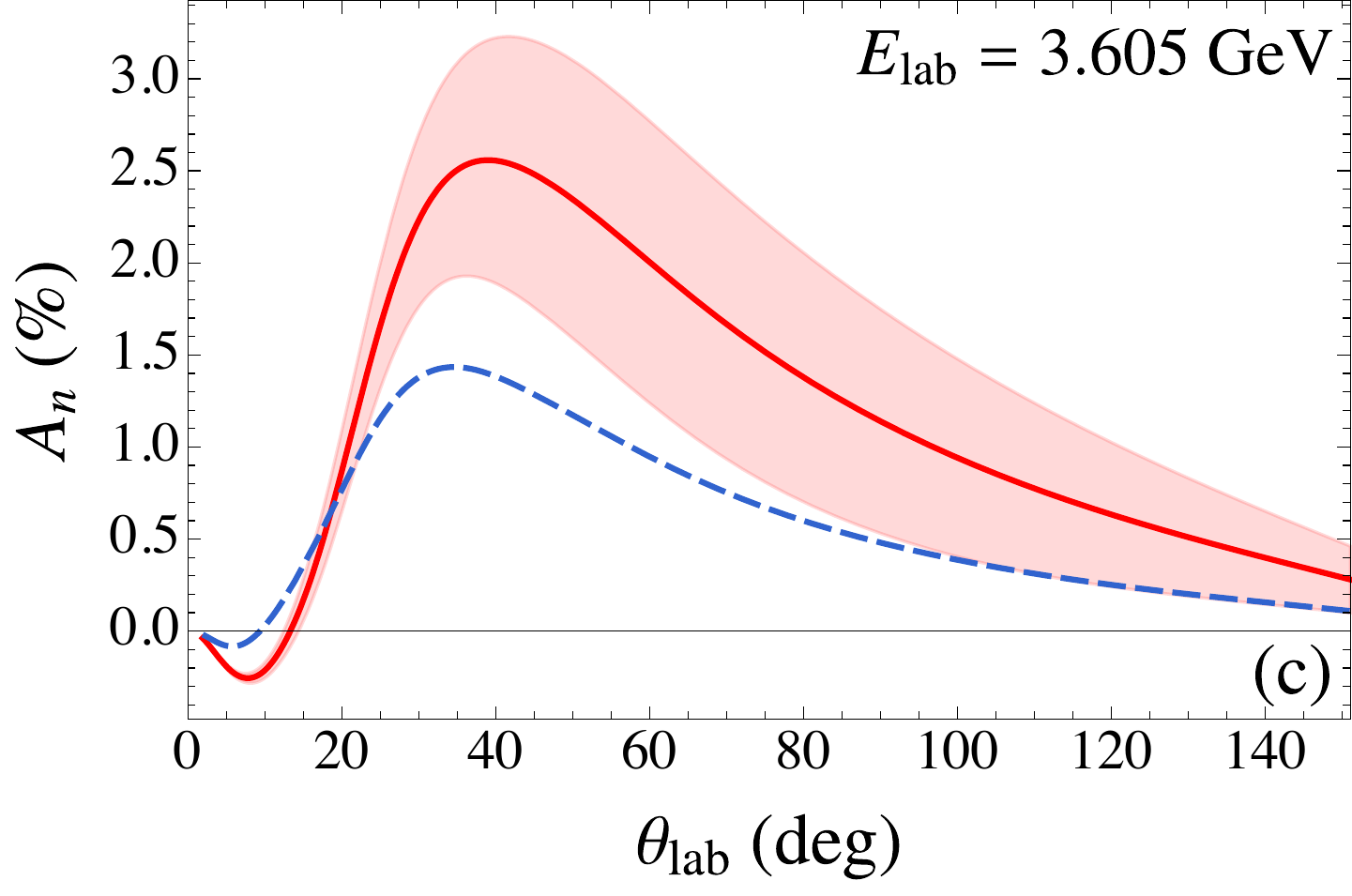}} \hfill
\subfloat{
\includegraphics[width=0.49\textwidth]{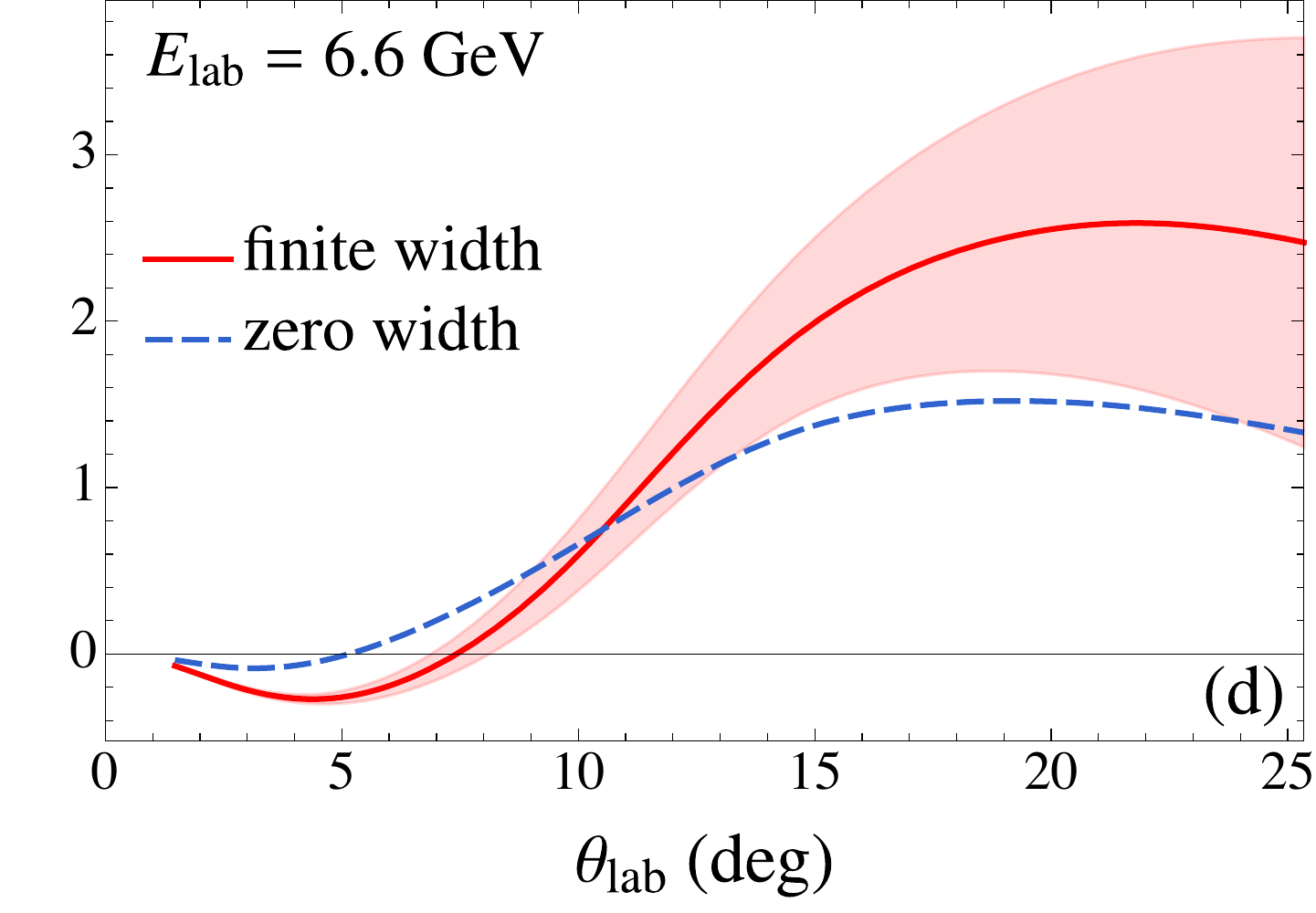}}
\caption{Total target normal SSA $A_n$, with the nucleon plus all nine resonance contributions, as a function of the scattering angle $\theta_{\textrm{lab}}$ at fixed beam energies $E_{\rm lab}$ of (a)~1.245~GeV, (b)~2.2~GeV, (c)~3.605~GeV, and (d)~6.6~GeV. The zero width results (blue dashed curves) are also shown for comparison.}
\label{fig.AnWidth}
\end{figure}

The total target normal SSA $A_n$, including contributions from the nucleon elastic and the nine spin 1/2 and 3/2 resonances, is illustrated in Fig.~\ref{fig.AnWidth} as a function of the scattering angle $\theta_{\rm lab}$ at the same four fixed beam energies.
The results of the finite width calculation, using a Sill distribution as in Eq.~(\ref{eq.BW}), are compared with the zero width approximation.
The finite width results are qualitatively similar to the approximated ones, but quantitatively there are clear differences in some kinematic regions.

In general, the zero width results have a smaller magnitude for the total $A_n$ than the finite width case.
However, as observed above, the net $A_n$ from the elastic nucleon and the resonances resembles the trend of the $\Delta(1232)$ state alone.
The overall magnitude of the asymmetry can also be well approximated by the sum of the elastic nucleon and $\Delta(1232)$ contributions.
As for the beam normal SSA $B_n$, contribution from higher spin states, with spin $\geq 5/2$, as well as nonresonant backgrounds may need to be considered in future.

Unfortunately, to date there have not been any direct measurements of $A_n$ in electron-proton scattering.
However, there has been a measurement of $A_n$ for electron scattering from polarized $^3$He in the quasi-elastic region at Jefferson Lab Hall~A \cite{Zhang:2015kna}, from which the electron-neutron asymmetry was extracted assuming an input $ep$ asymmetry.
The experiment scattered unpolarized electrons with energies $E_{\rm lab} = 1.245$, 2.425 and 3.605~GeV from a $^3$He target polarized normal to the scattering plane, with the scattered electrons detected at angle $\theta_{\rm lab} = 17^\circ$, corresponding to three different CM angles $\theta_{\rm cm} \simeq 32^\circ$, $41^\circ$, and $48^\circ$ for the three respective beam energies.
For the input proton SSA $A_n$, the elastic proton intermediate state contribution to the TPE amplitude from Ref.~\cite{Afanasev:2002gr}, giving $(0.01 \pm 0.22)\%$, $(0.24 \pm 2.96)\%$, and $(0.62 \pm 1.09)\%$ at the three beam energies, respectively, was used to extract the neutron asymmetry from the measured $^3$He SSA. 

\begin{table}[t]
\centering
\caption{Comparison of the target normal SSA $A_n$ calculated in this work with that used as input in the data analysis of electron--$^3$He scattering, at the beam energies $E_{\rm lab} = 1.245$, 2.425 and 3.605~GeV, and scattering angle $\theta_{\rm lab} = 17^\circ$~\cite{Zhang:2015kna}.}
\begin{tabular*}{\textwidth}{l @{\extracolsep{\fill}} ccc @{}}
\hline\hline
$E_{\rm lab}$ (GeV) & \multicolumn{2}{c}{$A_n$ (this work)} & {input $A_n$ in Ref.~\cite{Zhang:2015kna}~~~} \\
\cline{2-3}\cline{4-4}  
  & $N$ & $N$ + resonances & $N$ only\\
\hline 
~~~~1.245  &  0.008 & $-0.381 \pm 0.018$ & $0.01 \pm 0.22$ \\
~~~~2.425  &  0.173 & $-0.173 \pm 0.049$ & $0.24 \pm 2.96$ \\
~~~~3.605  &  0.400 & ~~$0.414 \pm 0.145$ & $0.62 \pm 1.09$ \\
\hline\hline
\end{tabular*}
\label{tab.An}
\end{table}

In contrast, in this work we find a total contribution to $A_n$ from the nucleon elastic state and the nine resonances of $(-0.381 \pm 0.018)\%, (-0.173 \pm 0.049)\%$, and $(0.414 \pm 0.145)\%$, at $\theta_{\rm lab} = 17^\circ$ and beam energies $E_{\rm lab} = 1.245$, 2.425 and 3.605~GeV, respectively.
Overall, the input proton asymmetry $A_n$ from Ref.~\cite{Afanasev:2002gr} is larger than our calculated result for the nucleon elastic state only, although consistent within the uncertainty.
The nucleon resonant contribution is sizeable at smaller beam energies, but is negligible at the highest energy, $E_{\rm lab} = 3.605$~GeV.  

% .............................................................................
\subsection{Beam transverse SSA $B_x$}
\label{ssec.BxResult}

A general electron spin vector transverse to the beam direction ($\bm{\hat z}$) is given by
\be
\bm{S} = \cos\phi_s\, \bm{\hat x} + \sin\phi_s\, \bm{\hat y},
\ee
where $\phi_s$ is the azimuthal angle with respect to the scattering ($\bm{\hat x} - \bm{\hat z}$) plane.
As discussed in Sec.~\ref{ssec.intro} above, interference between the OPE amplitude and the imaginary part of the TPE amplitude produces a beam normal SSA $B_n$ which depends only on the normal component of the spin.

In principle, a beam transverse SSA can also arise from the $x$-component of $\bm{S}$ due to a parity-violating interaction.
At lowest order this involves the interference of the OPE amplitude, ${\cal M}_\gamma$, and the $Z$-exchange amplitude, ${\cal M}_Z$.
This same interference gives the usual lowest order parity-violating asymmetry $A^\textrm{PV}$ for a longitudinally polarized beam,
\be
A^\textrm{PV} = -\frac{G_F}{\sqrt{2}} \frac{Q^2}{4\pi\alpha} \frac{1}{\sigma_R}\Bigl\{
g_V^e G_M G_A^Z\, \nu (1-\epsilon) - g_A^e \left[G_E G_E^Z\, \epsilon + G_M G_M^Z\, \tau \right]\Bigr\},
\label{eq.APV}
\ee
where $g_V^e$ and $g_A^e$ are the vector and axial-vector $eZ$ couplings, $G_A^Z$, $G_E^Z$,
and $G_M^Z$ are the proton weak form factors, $G_F$ is the Fermi constant, and the reduced cross section is $\sigma_R = G_E^2\, \epsilon + G_M^2\, \tau$.
The kinematic variables in Eq.~(\ref{eq.APV}) are dimensionless quantities that can be expressed in terms of the Mandelstam variables $s$, $t$, and $u$ as
\be
\tau = -\frac{t}{4 M^2},\qquad \nu = \frac{s-u}{4 M^2},\qquad
\epsilon = \frac{\nu^2-\tau(\tau+1)}{\nu^2+\tau(\tau+1)}.
\ee
The SSA for a purely transverse in-plane beam, denoted $B_x$, is given by
\bea
B_x &=& -\frac{G_F}{\sqrt{2}} \frac{Q^2}{4\pi\alpha} \frac{1}{\sigma_R} \frac{m_e}{E_\textrm{lab}}
\sqrt{\frac{2\epsilon(1-\epsilon)}{\tau+1}} \Bigl\{
g_V^e G_M G_M^Z\, 2 \tau (\tau+1)
\nonumber\\
&&\quad-\, g_A^e \left[G_E\, G_E^Z\, (\nu + \tau + 1) + G_M G_M^Z\, \tau (\nu - \tau - 1)\right]\Bigr\}.
\eea
In combination with $B_n$, this results in a general beam asymmetry of the form
\be
B_n \sin{\phi_s} + B_x \cos{\phi_s} = \sqrt{B_n^2+B_x^2}\, \sin({\phi_s + \delta_s});\qquad \delta_s = \tan^{-1}\left(\frac{B_x}{B_n}\right).
\ee
The general beam asymmetry then tretains a sinusoidal dependence on $\phi_s$, but with a phase shift $\delta_s$ relative to the pure beam normal SSA.

To obtain an order of magnitude estimate of the various asymmetries, we can write
\begin{subequations}
\bea
A^\textrm{PV} &\sim& \frac{Q^2}{M_Z^2} \approx Q^2 \times 10^{-4},\\
B_n &\sim& \alpha \frac{m_e}{M} \approx 5\times 10^{-6},\\
B_x &\sim& \frac{Q^2}{M_Z^2} \frac{m_e}{M} \approx Q^2\times \left(5\times 10^{-8}\right),
\eea
\end{subequations}
with $Q^2$ in units of GeV$^2$.
Aside from using muons instead of electrons, there seems to be no natural way to enhance the ratio $B_x/A^\textrm{PV}$ over the naive estimate of $5\times 10^{-4}$.
For the kinematics given in Table~\ref{tab.BnThExp}, the largest value of this ratio is $2\times 10^{-4}$ for the A4 (2020) kinematics, suggesting that the transverse parity-violating asymmetry is indeed negligible compared to the longitudinal asymmetry.
Measuring a phase shift $\delta_s \approx B_x/B_n$ seems equally unlikely, although the ratio could potentially be enhanced at higher $Q^2$.

%%%%%%%%%%%%%%%%%%%%%%%%%%%%%%%%%%%%%%%%%%%%%%%%%%%%%%%%%%%%%%%%%%%%%%%%%%%%%%%%%%
\section{Conclusions}
\label{sec.conclusions}

In this study we have calculated beam and target normal single-spin asymmetries in elastic electron-proton scattering using the imaginary part of two-photon exchange amplitudes, including contributions from $J^P = 1/2^\pm$ and $3/2^\pm$ excited state resonances with mass below 1.8~GeV.
For the resonance electrocouplings at the hadronic vertices we employed helicity amplitudes from the latest analysis of CLAS meson electroproduction data at $Q^2 \lesssim 5$~GeV$^2$. % \cite{HillerBlin:2019jgp}.

The effect of finite resonance widths on the beam normal SSA $B_n$ has been carefully investigated and found to be negligible in the forward angle region, becoming more noticeable at larger scattering angles.
We believe this may be attributable to a non-negligible contribution from the QRCS region above the nominal threshold excitation energy.

Among the various intermediate state contributions to $B_n$, the elastic nucleon and spin 1/2 resonances are suppressed by an order of magnitude or more compared to the spin 3/2 resonances.
The $\Delta(1232)$ resonance alone is a good approximation at forward angles for all beam energies.
The $N(1520)$ contribution is noticeably smaller than the $\Delta(1232)$, but both are negative across the range of energies and angles considered.
The $\Delta(1700)$ and $N(1720)$ are major contributors in the far forward and backward angle regions above their threshold excitation energies, both having positive contributions across energy and angle.
As a result, the total $B_n$ is somewhat sensitive to cancellations between the resonance contributions, changing from negative to positive with increasing energy and angle.
Uncertainties in the input electrocouplings are also significant for the $N(1520)$, $\Delta(1700)$ and $N(1720)$ states, leading to a rather large overall uncertainty band in the total $B_n$.

The results given in this work tend to overshoot the experimental $B_n$ data at lower beam energies $E_{\rm lab}<1$~GeV at both forward and backward angles.
This is the region in which the $\Delta(1232)$ dominates, with relatively small uncertainties in its input parameters.
There is good agreement between theory and the high-precision Qweak measurement at $E_{\rm lab}=1.149$~GeV, and modest agreement at the highest available energy $E_{\rm lab}\sim 3$~GeV and very forward angles, where the experimental uncertainties from the G0 and HAPPEX data are rather large.

For the target normal SSA $A_n$, the higher resonances beyond the $\Delta(1232)$ have almost no net effect.
Unlike $B_n$, the elastic nucleon intermediate state makes a significant contribution over the entire range of energy, $E_{\rm lab} \simeq 0.5$ to 6.6~GeV, considered in this work.
The sum of nucleon and $\Delta(1232)$ contributions account for most of the total $A_n$.
The spin 3/2 state $N(1520)$ is less significant for $A_n$ than it is for $B_n$, but the spin 1/2 state $N(1535)$ becomes a major contributor.
Also, unlike $B_n$, the peak magnitude of $A_n$ versus $\theta_{\rm lab}$ increases with energy in the range from $E_{\rm lab} = 0.5$ to 6.6~GeV.

For future work, given the significant uncertainties in the parameters of the higher mass resonances, better data to constrain electrocouplings for the higher mass excitations, such as the $\Delta(1700)$, would be helpful.
Effects of higher spin states, with spin $\geq 5/2$, can also be investigated, although uncertainties in the electrocouplings would limit the predictive power of such calculations.
Carlson {\it et al.}~\cite{Carlson:2017lys} also extended the calculation of beam SSAs from excited state resonance contributions to inelastic channels, such as the $ep \to e \Delta(1232)$ production process.
Finally, we note that an interesting quark level study~\cite{Gorchtein:2004ac} of beam normal SSAs, applicable at high-$Q^2 \gg M^2$ region, was performed in terms of a convolution of quark amplitudes and generalized parton distributions, which could be viewed as complementary to the resonance dominated region discussed in our analysis. 

%%%%%%%%%%%%%%%%%%%%%%%%%%%%%%%%%%%%%%%%%%%%%%%%%%%%%%%%%%%%
\acknowledgments

We thank Pratik Sachdeva for collaboration in the very early stages of this work, with support from
the DOE Science Undergraduate Laboratory Internship program.
This work was supported by the Natural Sciences and Engineering Research Council of Canada, and the US Department of Energy contract DE-AC05-06OR23177, under which Jefferson Science Associates, LLC operates Jefferson Lab. J.A. acknowledges the support from Shahjalal University of Science and Technology Research Centre.

%%%%%%%%%%%%%%%%%%%%%%%%%%%%%%%%%%%%%%%%%%%%%%%%%%%%%%%%%%%%
\appendix
\section{Numerical evaluation of $B_n$ in the QRCS region}
\label{sec.apndx}

We elaborate here on our semi-analytic method of evaluating the integral of Eq.~(\ref{eq.SSA6}) in the QRCS region $W\sim W_\textrm{max}=\sqrt{s}-m_e$. We define ${\cal B}_n(W)$ via
\be
B_n = \int_{W_\textrm{th}}^{W_\textrm{max}} \dd W\,{\cal B}_n (W),
\label{eq:BnW}
\ee
so that ${\cal B}_n(W)$ includes the angular integrals of Eq.~(\ref{eq.SSA6}). As discussed in Sec.~\ref{ssec:QRCS}, in the QRCS region the slowly varying tensor product $L_{\rho\mu\nu} H^{\rho\mu\nu}$ for $B_n$ in Eq.~(\ref{eq.SSA6}) is evaluated at
$Q_1^2=Q_2^2=0$, leaving a numerator independent of $\theta_{k_1}$ and $\phi_{k_1}$. The resulting expression is proportional to $J(W)$ as defined in Eq.~(\ref{eq.JW}), which can be evaluated analytically.
Applying Eqs.~(36-37) of Ref.~\cite{Blunden:2017nby} to the present case, we find
in agreement with Ref.~\cite{Gorchtein:2004ac}, that
\bea
J(W) =
\frac{\pi}{2 \sqrt{s}\, Q^2 E_k x_1}  \log\left(\frac{x_1+x_2}{x_1-x_2}\right),\quad
x_1=\sqrt{x_2^2+\frac{4 m_e^2}{Q^2} (1-z)^2},\quad x_2=\frac{|\bm{k}_1|}{E_k},
\label{eq:JWnum}
\eea
where $z=E_{k_1}/E_k$.

\begin{figure}[ht!]%
\graphicspath{{ImagesK/}}
\subfloat{
\includegraphics[width=0.49\textwidth]{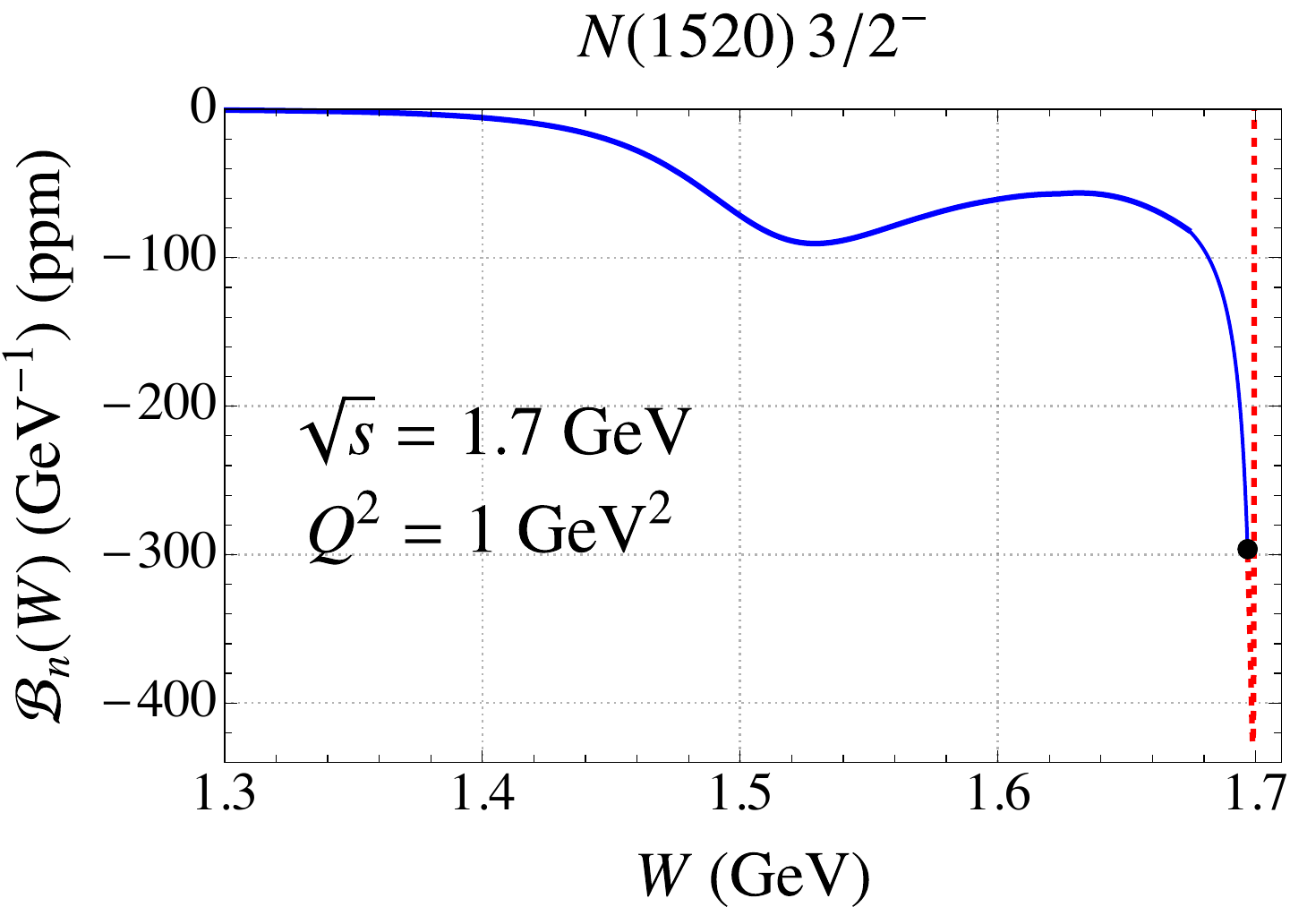}}
\hfill
\subfloat{
\includegraphics[width=0.49\textwidth]{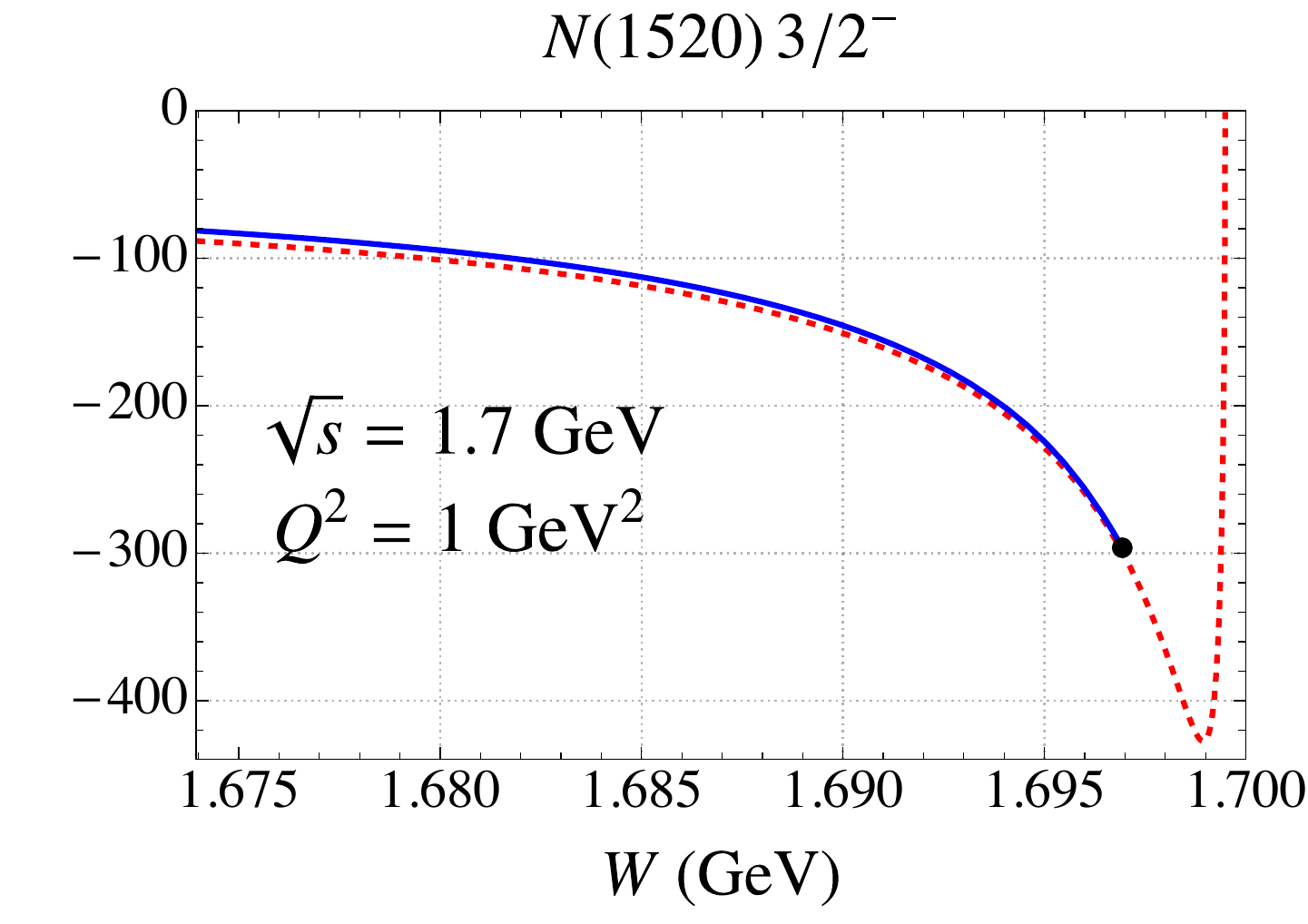}}
\caption{
The integrand ${\cal B}_n(W)$ of Eq.~(\ref{eq:BnW}) as a function of $W$ for the $N(1520)$ resonance at the sample kinematics $\sqrt{s}=1.7$~GeV and $Q^2=1$~GeV$^2$. The right panel is magnified to show the quasi-singular behaviour of ${\cal B}_n(W)$ as $W\to W_\textrm{max}=\sqrt{s}-m_e$.
The dashed red line makes use of the analytic result of Eq.~(\ref{eq:JWnum}), while the solid blue line is the fully numerical evaluation. The dot indicates our chosen matching point at $W=W_\textrm{max}-5 m_e$.}
\label{fig.QRCS}
\end{figure}

Figure~\ref{fig.QRCS} shows ${\cal B}_n(W)$ for the $N(1520)$ resonance at the sample kinematics of $\sqrt{s}=1.7~\textrm{GeV}$ 
($E_\textrm{lab}=1.071$~GeV) and $Q^2=1~\textrm{GeV}^2$.
This is above the nominal threshold energy of $E_\textrm{lab}^\textrm{th}=0.75$~GeV for excitation of a zero width resonance (see Table~\ref{tab.AhUncr}).
As shown in the left panel of Fig.~\ref{fig.QRCS}, due to the behaviour of $J(W)$, ${\cal B}_n(W)$ increases in magnitude with $W$ above threshold, and has an extremum near $W=W_\textrm{max}- m_e$ before falling sharply to 0 at $W=W_\textrm{max}$.
The right panel is magnified to show the matching between the full numerical and semi-analytical regions.
The dot indicates our chosen matching point at $W=W_\textrm{max}-5 m_e$.

%%%%%%%%%%%%%%%%%%%%%%%%%%%%%%%%%%%%%%%%%%%%%%%%%%%%%%%%%%%%
\newpage
\bibliography{SSA}

\end{document}